%% file: paper.tex
\documentclass{article}
\usepackage{amsfonts}
\usepackage{graphicx}
\usepackage{xspace}
\usepackage{hyperref}
\usepackage{subcaption}
\usepackage{amsmath}
\usepackage{amssymb}
\usepackage{graphicx}
\usepackage{program}
\usepackage{ifthen}
\usepackage{epsfig}
\usepackage{array}
\usepackage{epic}
\usepackage{color}
\usepackage[table]{xcolor}
\usepackage{hyperref}
\usepackage{xspace}
\usepackage{listings}
\usepackage{soul}
\usepackage{float}

\definecolor{mygray}{rgb}{0.9,0.9,0.9}
\definecolor{mymauve}{rgb}{0.58,0,0.82}
\definecolor{myred}{rgb}{0.72,0.18,0.0} 
\definecolor{mygreen}{rgb}    {0.0,0.72,0.0} 
\definecolor{myblue}{rgb} {0.18,0.0,0.72} 
\definecolor{mycreme}{rgb}        {1.0,0.8,0.2} 

\input{color_flatex}

\input{macros}

\newcommand{\pf}{{\tt PF}}
\newcommand{\tu}{{\tt TU}}
\newcommand{\pu}{{\tt PU}}

\newcommand{\gemm}{\mbox{\sc gemm}\xspace}

\newcommand{\bs}{b}

\title{Programming Parallel Dense Matrix Factorizations with Look-Ahead and OpenMP}
\author{%
Sandra Catal\'an%
\footnote{Depto. Ingenier\'{\i}a y Ciencia de Computadores, Universidad Jaume I, Castell\'on, Spain. {\tt \{catalans,adcastel,quintana\}@icc.uji.es}}
\and
Adri\'an Castell\'o$^*$
\and
Francisco D. Igual%
\footnote{Depto. de Arquitectura de Computadores y Autom\'atica, Universidad Complutense de Madrid, Spain 
          {\tt \{figual,rafaelrs\}@ucm.es}}
\and
Rafael Rodr\'{\i}guez-S\'anchez$^\dag$
\and
Enrique S. Quintana-Ort\'{\i}$^*$
}

\date{\today}

\begin{document}

\maketitle

\begin{abstract}
\input{s0-abstract}
\end{abstract}

\input{body}

\section*{Acknowledgments}

\input{sn-acks}

\bibliographystyle{plain}
\bibliography{biblio}

\end{document}

%% file: color_flatex.tex
\usepackage{arydshln}

\newcolumntype{I}{!{\vrule width 1.5pt}}

\newlength\savedwidth 

\newcommand\whline{\noalign{\global\savedwidth\arrayrulewidth 
                            \global\arrayrulewidth 1.5pt}%
           \hline 
           \noalign{\global\arrayrulewidth\savedwidth}}

\newcommand{\FlaTwoByTwo}[4]{
\left(
\begin{array}{c I c}
#1 & #2 \\ \whline 
#3 & #4 
\end{array}
\right) 
}

\newcommand{\FlaThreeByThreeTL}[9]{
\left(
\begin{array}{c | c I c}
#1 & #2 & #3 \\ \hline
#4 & #5 & #6 \\ \whline
#7 & #8 & #9
\end{array}
\right)
}

\newcommand{\FlaThreeByThreeBR}[9]{
\left(
\begin{array}{c I c | c}
#1 & #2 & #3 \\ \whline 
#4 & #5 & #6 \\ \hline 
#7 & #8 & #9 
\end{array}
\right) 
}

\newcommand{\operation}{}

\newcommand{\routinename}{}

\newcommand{\precondition}{~}

\newcommand{\postcondition}{~}

\newcommand{\invariant}{~}

\newcommand{\guard}{~}

\newcommand{\partitionings}{~}

\newcommand{\partitionsizes}{~}

\newcommand{\blocksize}{blank}

\newcommand{\repartitionings}{~}

\newcommand{\repartitionsizes}{~}

\newcommand{\moveboundaries}{~}

\newcommand{\beforeupdate}{~}

\newcommand{\afterupdate}{~}

\newcommand{\update}{~}

\newcommand{\resetsteps}{

\renewcommand{\operation}{\phantom{[A] = op( A )}}

\renewcommand{\routinename}{\operation}

\renewcommand{\precondition}{\phantom{A = \widehat A}}

\renewcommand{\postcondition}{\phantom{A = \widehat A}}

\renewcommand{\invariant}{\phantom{ \FlaTwoByTwo{A_{TL}}{A_{TR}}{A_{BL}}{A_{BR}} =
		\FlaTwoByTwo{A_{TL}}{A_{TR}}{A_{BL}}{A_{BR}}
		\wedge
		\FlaTwoByTwo{A_{TL}}{A_{TR}}{A_{BL}}{A_{BR}} =
		\FlaTwoByTwo{A_{TL}}{A_{TR}}{A_{BL}}{A_{BR}}~~~~~~~
		}}

\renewcommand{\blocksize}{blank}

\renewcommand{\guard}{\phantom{m( A_{BL} ) < m( A )}}

\renewcommand{\partitionings}{
$
\phantom{\FlaTwoByTwo{A_{TL}}{A_{TR}}{A_{BL}}{A_{BR}}
\rightarrow
\FlaThreeByThreeBR
   {A_{00}}{a_{01}}{A_{02}}
   {a_{10}^T}{\alpha_{11}}{a_{12}^T}
   {A_{20}}{a_{21}}{A_{22}}}   
$
}

\renewcommand{\partitionsizes}{$ \phantom{m( A )} $}

\renewcommand{\repartitionings}{
$
\phantom{\FlaTwoByTwo{A_{TL}}{A_{TR}}{A_{BL}}{A_{BR}}
\rightarrow
\FlaThreeByThreeBR
   {A_{00}}{a_{01}}{A_{02}}
   {a_{10}^T}{\alpha_{11}}{a_{12}^T}
   {A_{20}}{a_{21}}{A_{22}}}   
$
}

\renewcommand{\repartitionsizes}{$\phantom{m(A)}$}

\renewcommand{\moveboundaries}{
$
\phantom{\FlaTwoByTwo{A_{TL}}{A_{TR}}{A_{BL}}{A_{BR}}
\rightarrow
\FlaThreeByThreeBR
   {A_{00}}{a_{01}}{A_{02}}
   {a_{10}^T}{\alpha_{11}}{a_{12}^T}
   {A_{20}}{a_{21}}{A_{22}}}   
$
}

\renewcommand{\beforeupdate}{
\phantom{\FlaTwoByTwo{A_{TL}}{A_{TR}}{A_{BL}}{A_{BR}}
\rightarrow
\FlaThreeByThreeBR
   {A_{00}}{a_{01}}{A_{02}}
   {a_{10}^T}{\alpha_{11}}{a_{12}^T}
   {A_{20}}{a_{21}}{A_{22}}}   
}

\renewcommand{\afterupdate}{
\phantom{\FlaTwoByTwo{A_{TL}}{A_{TR}}{A_{BL}}{A_{BR}}
\rightarrow
\FlaThreeByThreeBR
   {A_{00}}{a_{01}}{A_{02}}
   {a_{10}^T}{\alpha_{11}}{a_{12}^T}
   {A_{20}}{a_{21}}{A_{22}}}   
}

\renewcommand{\update}{
\phantom{$
\begin{array}{l}
\\
\\
\\
\end{array}
$}
}
}

\newcommand{\NoShow}[1]{}

\newcommand{\FlaAlgorithm}{
\begin{tabular}{| p{0.92\textwidth}|} \hline
$\mbox{\color{blue}Algorithm:~}\routinename$
\\ \whline
\partitionings \\
$\mbox{\color{blue} ~~~where~}$ \partitionsizes 
\\ 
$\mbox{\color{blue}while~} \guard \mbox{~\color{blue} do}$
\\
\ifthenelse{\equal{\blocksize}{1}}{\\}%
{%
\ifthenelse{ \equal{\blocksize}{blank} }{}%
{~~~~{\bf Determine block size $ \blocksize $}\\}%
}
~~~~ 
\repartitionings \\
~~~$\mbox{\color{blue} ~~~where~}$ \repartitionsizes
\\ \hline
~~~~  \update 
\\ \hline
~~~~ 
\moveboundaries 
\\
$\mbox{\color{blue} endwhile} $
\\ \hline 
\end{tabular}
}

\newcommand{\FlaWorksheet}{
\begin{tabular}{| c | p{0.98\textwidth} |}\hline
Step & $\mbox{\color{blue}Algorithm:~}\routinename$
\\ \hline
\rowcolor{yellow!75}
1a & $ \precondition $ 
\\ \whline
4 & 
\begin{minipage}[t]{0.9\textwidth}%
\partitionings~ \\
$\mbox{\color{blue} ~~~where~}$ \partitionsizes
\end{minipage}
\\ \hline
\rowcolor{yellow!75}
2 & $ \invariant $ 
\\ \hline
3 &$\mbox{\color{blue}while~} \guard \mbox{~\color{blue} do}$
\\ \hline 
\rowcolor{yellow!75}
2,3 & ~~~~ $ \invariant \wedge \guard$ 
\\ \hline
5a & ~~~~ \begin{minipage}[t]{0.85\textwidth}%
\ifthenelse{\equal{\blocksize}{1}}{}%
{%
\ifthenelse{ \equal{\blocksize}{blank} }{}%
{{\bf Determine block size $ \blocksize $}\\}%
}
\repartitionings~ \\
$\mbox{\color{blue} ~~~where~}$ \repartitionsizes
\end{minipage}
\\ \hline
\rowcolor{yellow!75}
6 & ~~~~ $\beforeupdate $
\\ \hline
8 & ~~~~  \update 
\\ \hline
5b & ~~~~ \begin{minipage}[t]{0.85\textwidth}%
\moveboundaries~
\end{minipage}
\\ \hline
\rowcolor{yellow!75}
7 & ~~~~ $\afterupdate $
\\ \hline
\rowcolor{yellow!75}
2 & ~~~~ $ \invariant  $ 
\\ \hline
 &$\mbox{\color{blue} endwhile} $
\\ \hline \whline
\rowcolor{yellow!75}
2,3 & $ \invariant \wedge \neg( \guard )$ 
\\ \hline
\rowcolor{yellow!75}
1b & $ \postcondition $ 
\\ \hline
\end{tabular}
}

\newcommand{\FlaWorksheetNine}{
\begin{tabular}{| c | p{0.98\textwidth} |}\hline
Step & $\mbox{\color{blue}Algorithm:~}\routinename$
\\ \hline
\rowcolor{yellow!75}
\phantom{1a} & $ \phantom\precondition $ 
\\ \whline
\phantom{4} & 
\begin{minipage}[t]{0.9\textwidth}%
\partitionings~ \\
$\mbox{\color{blue} ~~~where~}$ \partitionsizes
\end{minipage}
\\ \hline
\rowcolor{yellow!75}
\phantom{2} & $ \phantom\invariant $ 
\\ \hline
\phantom{3} &$\mbox{\color{blue}while~} \guard \mbox{~\color{blue} do}$
\\ \hline 
\rowcolor{yellow!75}
\phantom{2,3} & ~~~~ $ \phantom\invariant \phantom \wedge \phantom
                \guard $ 
\\ \hline
\phantom{5a} & ~~~~ \begin{minipage}[t]{0.85\textwidth}%
\ifthenelse{\equal{\blocksize}{1}}{}%
{%
\ifthenelse{ \equal{\blocksize}{blank} }{}%
{{\bf Determine block size $ \blocksize $}\\}%
}
\repartitionings~ \\
$\mbox{\color{blue} ~~~where~}$ \repartitionsizes
\end{minipage}
\\ \hline
\rowcolor{yellow!75}
\phantom{6} & ~~~~ $\phantom\beforeupdate $
\\ \hline
\phantom{8} & ~~~~  \update 
\\ \hline
\phantom{5b} & ~~~~ \begin{minipage}[t]{0.85\textwidth}%
\moveboundaries~
\end{minipage}
\\ \hline
\rowcolor{yellow!75}
\phantom{7} & ~~~~ $\phantom\afterupdate $
\\ \hline
\rowcolor{yellow!75}
\phantom{2} & ~~~~ $ \phantom\invariant  $ 
\\ \hline
 &$\mbox{\color{blue} endwhile} $
\\ \hline \whline
\rowcolor{yellow!75}
\phantom{2,3} & $ \phantom\invariant \wedge \neg( \phantom\guard )$ 
\\ \hline
\rowcolor{yellow!75}
\phantom{1b} & $ \phantom\postcondition $ 
\\ \hline
\end{tabular}
}

\newcommand{\FlaWorksheetEight}{
\begin{tabular}{| c | p{0.98\textwidth} |}\hline
Step & $\mbox{\color{blue}Algorithm:~}\routinename$
\\ \hline
\rowcolor{yellow!75}
1a & $ \precondition $ 
\\ \whline
4 & 
\begin{minipage}[t]{0.9\textwidth}%
\partitionings~ \\
$\mbox{\color{blue} ~~~where~}$ \partitionsizes
\end{minipage}
\\ \hline
\rowcolor{yellow!75}
2 & $ \invariant $ 
\\ \hline
3 &$\mbox{\color{blue}while~} \guard \mbox{~\color{blue} do}$
\\ \hline 
\rowcolor{yellow!75}
2,3 & ~~~~ $ \invariant \wedge \guard $ 
\\ \hline
5a & ~~~~ \begin{minipage}[t]{0.85\textwidth}%
\ifthenelse{\equal{\blocksize}{1}}{}%
{%
\ifthenelse{ \equal{\blocksize}{blank} }{}%
{{\bf Determine block size $ \blocksize $}\\}%
}
\repartitionings~ \\
$\mbox{\color{blue} ~~~where~}$ \repartitionsizes
\end{minipage}
\\ \hline
\rowcolor{yellow!75}
6 & ~~~~ $\beforeupdate $
\\ \hline
\rowcolor{orange!50}    
8 & ~~~~  \update 
\\ \hline
5b & ~~~~ \begin{minipage}[t]{0.85\textwidth}%
\moveboundaries~
\end{minipage}
\\ \hline
\rowcolor{yellow!75}
7 & ~~~~ $\afterupdate $
\\ \hline
\rowcolor{yellow!75}
2 & ~~~~ $ \invariant  $ 
\\ \hline
 &$\mbox{\color{blue} endwhile} $
\\ \hline \whline
\rowcolor{yellow!75}
2,3 & $ \invariant \wedge \neg( \guard )$ 
\\ \hline
\rowcolor{yellow!75}
1b & $ \postcondition $ 
\\ \hline
\end{tabular}
}

\newcommand{\FlaWorksheetSeven}{
\begin{tabular}{| c | p{0.98\textwidth} |}\hline
Step & $\mbox{\color{blue}Algorithm:~}\routinename$
\\ \hline
\rowcolor{yellow!75}
1a & $ \precondition $ 
\\ \whline
4 & 
\begin{minipage}[t]{0.9\textwidth}%
\partitionings~ \\
$\mbox{\color{blue} ~~~where~}$ \partitionsizes
\end{minipage}
\\ \hline
\rowcolor{yellow!75}
2 & $ \invariant $ 
\\ \hline
3 &$\mbox{\color{blue}while~} \guard \mbox{~\color{blue} do}$
\\ \hline 
\rowcolor{yellow!75}
2,3 & ~~~~ $ \invariant \wedge \guard$ 
\\ \hline
5a & ~~~~ \begin{minipage}[t]{0.85\textwidth}%
\ifthenelse{\equal{\blocksize}{1}}{}%
{%
\ifthenelse{ \equal{\blocksize}{blank} }{}%
{{\bf Determine block size $ \blocksize $}\\}%
}
\repartitionings~ \\
$\mbox{\color{blue} ~~~where~}$ \repartitionsizes
\end{minipage}
\\ \hline
\rowcolor{yellow!75}
6 & ~~~~ $\beforeupdate $
\\ \hline
8 & ~~~~  \phantom\update 
\\ \hline
5b & ~~~~ \begin{minipage}[t]{0.85\textwidth}%
\moveboundaries~
\end{minipage}
\\ \hline
\rowcolor{orange!50}    
7 & ~~~~ $\afterupdate $
\\ \hline
\rowcolor{yellow!75}
2 & ~~~~ $ \invariant  $ 
\\ \hline
 &$\mbox{\color{blue} endwhile} $
\\ \hline \whline
\rowcolor{yellow!75}
2 & $ \invariant \wedge \neg( \guard )$ 
\\ \hline
\rowcolor{yellow!75}
1b & $ \postcondition $ 
\\ \hline
\end{tabular}
}

\newcommand{\FlaWorksheetSix}{
\begin{tabular}{| c | p{0.98\textwidth} |}\hline
Step & $\mbox{\color{blue}Algorithm:~}\routinename$
\\ \hline
\rowcolor{yellow!75}
1a & $ \precondition $ 
\\ \whline
4 & 
\begin{minipage}[t]{0.9\textwidth}%
\partitionings~ \\
$\mbox{\color{blue} ~~~where~}$ \partitionsizes
\end{minipage}
\\ \hline
\rowcolor{yellow!75}
2 & $ \invariant $ 
\\ \hline
3 &$\mbox{\color{blue}while~} \guard \mbox{~\color{blue} do}$
\\ \hline 
\rowcolor{yellow!75}
2,3 & ~~~~ $ \invariant \wedge \guard $ 
\\ \hline
5a & ~~~~ \begin{minipage}[t]{0.85\textwidth}%
\ifthenelse{\equal{\blocksize}{1}}{}%
{%
\ifthenelse{ \equal{\blocksize}{blank} }{}%
{{\bf Determine block size $ \blocksize $}\\}%
}
\repartitionings~ \\
$\mbox{\color{blue} ~~~where~}$ \repartitionsizes
\end{minipage}
\\ \hline
\rowcolor{orange!50}   
6 & ~~~~ $\beforeupdate $
\\ \hline
8 & ~~~~  \phantom\update 
\\ \hline
5b & ~~~~ \begin{minipage}[t]{0.85\textwidth}%
\moveboundaries~
\end{minipage}
\\ \hline
\rowcolor{yellow!75}
7 & ~~~~ $\phantom\afterupdate $
\\ \hline
\rowcolor{yellow!75}
2 & ~~~~ $ \invariant  $ 
\\ \hline
 &$\mbox{\color{blue} endwhile} $
\\ \hline \whline
\rowcolor{yellow!75}
2,3 & $ \invariant \wedge \neg( \guard )$ 
\\ \hline
\rowcolor{yellow!75}
1b & $ \postcondition $ 
\\ \hline
\end{tabular}
}

\newcommand{\FlaWorksheetFive}{
\begin{tabular}{| c | p{0.98\textwidth} |}\hline
Step & $\mbox{\color{blue}Algorithm:~}\routinename$
\\ \hline
\rowcolor{yellow!75}
1a & $ \precondition $ 
\\ \whline
4 & 
\begin{minipage}[t]{0.9\textwidth}%
\partitionings~ \\
$\mbox{\color{blue} ~~~where~}$ \partitionsizes
\end{minipage}
\\ \hline
\rowcolor{yellow!75}
2 & $ \invariant $ 
\\ \hline
3 &$\mbox{\color{blue}while~} \guard \mbox{~\color{blue} do}$
\\ \hline 
\rowcolor{yellow!75}
2,3 & ~~~~ $ \invariant \wedge \guard $ 
\\ \hline
\rowcolor{orange!50}   
5a & ~~~~ \begin{minipage}[t]{0.85\textwidth}%
\ifthenelse{\equal{\blocksize}{1}}{}%
{%
\ifthenelse{ \equal{\blocksize}{blank} }{}%
{{\bf Determine block size $ \blocksize $}\\}%
}
\repartitionings~ \\
$\mbox{\color{blue} ~~~where~}$ \repartitionsizes
\end{minipage}
\\ \hline
\rowcolor{yellow!75}
6 & ~~~~ $\phantom\beforeupdate $
\\ \hline
8 & ~~~~  \phantom\update 
\\ \hline
\rowcolor{orange!50}   
5b & ~~~~ \begin{minipage}[t]{0.85\textwidth}%
\moveboundaries~
\end{minipage}
\\ \hline
\rowcolor{yellow!75}
7 & ~~~~ $\phantom\afterupdate $
\\ \hline
\rowcolor{yellow!75}
2 & ~~~~ $ \invariant  $ 
\\ \hline
 &$\mbox{\color{blue} endwhile} $
\\ \hline \whline
\rowcolor{yellow!75}
2,3 & $ \invariant \wedge \neg( \guard )$ 
\\ \hline
\rowcolor{yellow!75}
1b & $ \postcondition $ 
\\ \hline
\end{tabular}
}

\newcommand{\FlaWorksheetFour}{
\begin{tabular}{| c | p{0.98\textwidth} |}\hline
Step & $\mbox{\color{blue}Algorithm:~}\routinename$
\\ \hline
\rowcolor{yellow!75}
1a & $ \precondition $ 
\\ \whline
\rowcolor{orange!50}   
4 & 
\begin{minipage}[t]{0.9\textwidth}%
\partitionings~ \\
$\mbox{\color{blue} ~~~where~}$ \partitionsizes
\end{minipage}
\\ \hline
\rowcolor{yellow!75}
2 & $ \invariant $ 
\\ \hline
3 &$\mbox{\color{blue}while~} \guard \mbox{~\color{blue} do}$
\\ \hline 
\rowcolor{yellow!75}
2,3 & ~~~~ $ \invariant \wedge \guard $ 
\\ \hline
5a & ~~~~ \begin{minipage}[t]{0.85\textwidth}%
\ifthenelse{\equal{\blocksize}{1}}{}%
{%
\ifthenelse{ \equal{\blocksize}{blank} }{}%
{{\bf Determine block size $ \phantom\blocksize $}\\}%
}
$\mbox{\phantom\repartitionings}$~ \\
$\mbox{\color{blue} ~~~where~}$ \phantom\repartitionsizes
\end{minipage}
\\ \hline
\rowcolor{yellow!75}
6 & ~~~~ $\phantom\beforeupdate $
\\ \hline
8 & ~~~~  \phantom\update 
\\ \hline
5b & ~~~~ \begin{minipage}[t]{0.85\textwidth}%
\phantom\moveboundaries~
\end{minipage}
\\ \hline
\rowcolor{yellow!75}
7 & ~~~~ $\phantom\afterupdate $
\\ \hline
\rowcolor{yellow!75}
2 & ~~~~ $ \invariant  $ 
\\ \hline
 &$\mbox{\color{blue} endwhile} $
\\ \hline \whline
\rowcolor{yellow!75}
2,3 & $ \invariant \wedge \neg( \guard )$ 
\\ \hline
\rowcolor{yellow!75}
1b & $ \postcondition $ 
\\ \hline
\end{tabular}
}

\newcommand{\FlaWorksheetThree}{
\begin{tabular}{| c | p{0.98\textwidth} |}\hline
Step & $\mbox{\color{blue}Algorithm:~}\routinename$
\\ \hline
\rowcolor{yellow!75}
1a & $ \precondition $ 
\\ \whline
4 & 
\begin{minipage}[t]{0.9\textwidth}%
$\mbox{\phantom{\partitionings}}$~ \\
$\mbox{\color{blue} ~~~where~}$\phantom{\partitionsizes}  
\end{minipage}
\\ \hline
\rowcolor{yellow!75}
2 & $ \invariant $ 
\\ \hline
\rowcolor{orange!50}  
3 &$\mbox{\color{blue}while~} \guard \mbox{~\color{blue} do}$
\\ \hline 
\rowcolor{orange!50}   
2,3 & ~~~~ $ \invariant \wedge \guard $ 
\\ \hline
5a & ~~~~ \begin{minipage}[t]{0.85\textwidth}%
\ifthenelse{\equal{\blocksize}{1}}{}%
{%
\ifthenelse{ \equal{\blocksize}{blank} }{}%
{{\bf Determine block size $ \phantom\blocksize $}\\}%
}
$\mbox{\phantom\repartitionings}$~ \\
$\mbox{\color{blue} ~~~where~}$ \phantom\repartitionsizes
\end{minipage}
\\ \hline
\rowcolor{yellow!75}
6 & ~~~~ $\phantom\beforeupdate $
\\ \hline
8 & ~~~~  \phantom\update 
\\ \hline
5b & ~~~~ \begin{minipage}[t]{0.85\textwidth}%
\phantom\moveboundaries~
\end{minipage}
\\ \hline
\rowcolor{yellow!75}
7 & ~~~~ $\phantom\afterupdate $
\\ \hline
\rowcolor{yellow!75}
2 & ~~~~ $ \invariant  $ 
\\ \hline
 &$\mbox{\color{blue} endwhile} $
\\ \hline \whline
\rowcolor{orange!50}   
2,3 & $ \invariant \wedge \neg( \guard )$ 
\\ \hline
\rowcolor{yellow!75}
1b & $ \postcondition $ 
\\ \hline
\end{tabular}
}

\newcommand{\FlaWorksheetTwo}{
\begin{tabular}{| c | p{0.98\textwidth} |}\hline
Step & $\mbox{\color{blue}Algorithm:~}\routinename$
\\ \hline
\rowcolor{yellow!75}
1a & $ \precondition $ 
\\ \whline
4 & 
\begin{minipage}[t]{0.9\textwidth}%
$\mbox{\phantom{\partitionings}}$~ \\
$\mbox{\color{blue} ~~~where~} $ \phantom{\partitionsizes} 
\end{minipage}
\\ \hline
\rowcolor{orange!50} 
2 & $ \invariant $ 
\\ \hline
3 &$\mbox{\color{blue}while~} \phantom\guard \mbox{~\color{blue} do}$
\\ \hline 
\rowcolor{orange!50} 
2,3 & ~~~~ $ \invariant \wedge \phantom \guard $ 
\\ \hline
5a & ~~~~ \begin{minipage}[t]{0.85\textwidth}%
\ifthenelse{\equal{\blocksize}{1}}{}%
{%
\ifthenelse{ \equal{\blocksize}{blank} }{}%
{{\bf Determine block size $ \phantom\blocksize $}\\}%
}
$\mbox{\phantom\repartitionings}$~ \\
$\mbox{\color{blue} ~~~where~}$ \phantom\repartitionsizes
\end{minipage}
\\ \hline
\rowcolor{yellow!75}
6 & ~~~~ $\phantom\beforeupdate $
\\ \hline
8 & ~~~~  \phantom\update 
\\ \hline
5b & ~~~~ \begin{minipage}[t]{0.85\textwidth}%
\phantom\moveboundaries~
\end{minipage}
\\ \hline
\rowcolor{yellow!75}
7 & ~~~~ $\phantom\afterupdate $
\\ \hline
\rowcolor{orange!50} 
2 & ~~~~ $ \invariant  $ 
\\ \hline
 &$\mbox{\color{blue} endwhile} $
\\ \hline \whline
\rowcolor{orange!50} 
2 & $ \invariant \wedge \neg( \phantom\guard )$ 
\\ \hline
\rowcolor{yellow!75}
1b & $ \postcondition $ 
\\ \hline
\end{tabular}
}

\newcommand{\FlaWorksheetOne}{
\begin{tabular}{| c | p{0.98\textwidth} |}\hline
Step & $\mbox{\color{blue}Algorithm:~}\routinename$
\\ \hline
\rowcolor{orange!50}
1a & $ \precondition $ 
\\ \whline
4 & 
\begin{minipage}[t]{0.9\textwidth}%
$\mbox{\phantom{\partitionings}}$ ~ \\
$\mbox{\color{blue} ~~~where~}$ \phantom{\partitionsizes} 
\end{minipage}
\\ \hline
\rowcolor{yellow!75}
2 & $ \phantom\invariant $ 
\\ \hline
3 &$\mbox{\color{blue}while~} \phantom\guard \mbox{~\color{blue} do}$
\\ \hline 
\rowcolor{yellow!75}
2,3 & ~~~~ $ \phantom\invariant \wedge \phantom \guard$ 
\\ \hline
5a & ~~~~ \begin{minipage}[t]{0.85\textwidth}%
\ifthenelse{\equal{\blocksize}{1}}{}%
{%
\ifthenelse{ \equal{\blocksize}{blank} }{}%
{{\bf Determine block size $ \phantom\blocksize $}\\}%
}
$\mbox{\phantom\repartitionings}$ ~ \\
$\mbox{\color{blue} ~~~where~}$ \phantom\repartitionsizes
\end{minipage}
\\ \hline
\rowcolor{yellow!75}
6 & ~~~~ $\phantom\beforeupdate $
\\ \hline
8 & ~~~~  \phantom\update 
\\ \hline
5b & ~~~~ \begin{minipage}[t]{0.85\textwidth}%
\phantom\moveboundaries~
\end{minipage}
\\ \hline
\rowcolor{yellow!75}
7 & ~~~~ $\phantom\afterupdate $
\\ \hline
\rowcolor{yellow!75}
2 & ~~~~ $ \phantom\invariant  $ 
\\ \hline
 &$\mbox{\color{blue} endwhile} $
\\ \hline \whline
\rowcolor{yellow!75}
2,3 & $ \phantom\invariant \wedge \neg( \phantom\guard )$ 
\\ \hline
\rowcolor{orange!50}
1b & $ \postcondition $ 
\\ \hline
\end{tabular}
}

\newcommand{\FlaWorksheetZero}{
\begin{tabular}{| c | p{0.98\textwidth} |}\hline
Step & $\mbox{\color{blue}Algorithm:~}\routinename$
\\ \hline
\rowcolor{yellow!75}
1a & $ \phantom\precondition $ 
\\ \whline
4 & 
\begin{minipage}[t]{0.9\textwidth}%
$\mbox{\phantom{\partitionings}}$ ~ \\
$\mbox{\color{blue} ~~~where~}$ \phantom{\partitionsizes} 
\end{minipage}
\\ \hline
\rowcolor{yellow!75}
2 & $ \phantom\invariant $ 
\\ \hline
3 &$\mbox{\color{blue}while~} \phantom\guard \mbox{~\color{blue} do}$
\\ \hline 
\rowcolor{yellow!75}
2,3 & ~~~~ $ \phantom\invariant \wedge \phantom \guard$ 
\\ \hline
5a & ~~~~ \begin{minipage}[t]{0.85\textwidth}%
\ifthenelse{\equal{\blocksize}{1}}{}%
{%
\ifthenelse{ \equal{\blocksize}{blank} }{}%
{{\bf Determine block size $ \phantom\blocksize $}\\}%
}
$\mbox{\phantom\repartitionings}$~ \\
$\mbox{\color{blue} ~~~where~}$ \phantom\repartitionsizes
\end{minipage}
\\ \hline
\rowcolor{yellow!75}
6 & ~~~~ $\phantom\beforeupdate $
\\ \hline
8 & ~~~~  \phantom\update 
\\ \hline
5b & ~~~~ \begin{minipage}[t]{0.85\textwidth}%
\phantom\moveboundaries~
\end{minipage}
\\ \hline
\rowcolor{yellow!75}
7 & ~~~~ $\phantom\afterupdate $
\\ \hline
\rowcolor{yellow!75}
2 & ~~~~ $ \phantom\invariant  $ 
\\ \hline
 &$\mbox{\color{blue} endwhile} $
\\ \hline \whline
\rowcolor{yellow!75}
2,3 & $ \phantom\invariant \wedge \neg( \phantom\guard )$ 
\\ \hline
\rowcolor{yellow!75}
1b & $ \postcondition $ 
\\ \hline
\end{tabular}
}

\newcommand{\TBTinitialize}{}
\newcommand{\FlaAlgorithmTBT}{
\begin{tabular}{|l|} \hline
$\mbox{\color{blue}Algorithm:~}\routinename$
\\ \whline
\partitionings \\
$\mbox{\color{blue} ~~~where~}$ \partitionsizes 
\\ 
\TBTinitialize\\
$\mbox{\color{blue}while~} \guard \mbox{~\color{blue} do}$
\\
\ifthenelse{\equal{\blocksize}{1}}{}%
{%
\ifthenelse{ \equal{\blocksize}{blank} }{}%
{~~~~{\bf Determine block size $ \blocksize $}\\}%
}
~~~~ 
\repartitionings \\
~~~$\mbox{\color{blue} ~~~where~}$ \repartitionsizes
\\ \hline
~~~~  \update 
\\ \hline
~~~~ 
\moveboundaries 
\\
$\mbox{\color{blue} endwhile} $
\\ \hline 
\end{tabular}
}

%% file: macros.tex
\newcommand{\trilu}[1]{\mbox{\sc trilu}( #1 )}

%% file: s0-abstract.tex
We investigate a parallelization strategy for dense matrix factorization (DMF) algorithms, using OpenMP, that departs from the 
legacy (or conventional) solution, which simply extracts concurrency from a multi-threaded version of BLAS. 
This approach is also different from the more sophisticated
runtime-assisted implementations, which decompose the operation into tasks and identify dependencies via directives and
runtime support.
Instead, our strategy attains high performance by explicitly embedding a static look-ahead technique into the 
DMF code, in order to overcome the performance bottleneck of the panel factorization, 
and realizing the trailing update via a cache-aware multi-threaded implementation of the BLAS. 
Although the parallel algorithms are specified with a high-level of abstraction,
the actual implementation can be easily derived from them, paving the road to deriving
a high performance implementation of a considerable fraction of LAPACK functionality on any multicore
platform with an OpenMP-like runtime.

%% file: body.tex
\input{s1-intro}

\input{s2-gemm}

\input{s3-dmf}
\input{s4-openmp}
\input{s5-lwt}
\input{s6-experiments}
\input{s7-remarks}

%% file: s1-intro.tex
\section{Introduction}

Dense linear algebra (DLA) lies at the bottom of the ``food chain'' for many scientific and engineering applications,
which require numerical kernels
to tackle linear systems, linear least squares problems or eigenvalue computations, among other problems~\cite{Dem97}.
In response, the scientific community has created
the Basic Linear Algebra Subroutines (BLAS) and the Linear Algebra Package (LAPACK)~\cite{BLAS3,lapack}.
These libraries standardize domain-specific interfaces for DLA operations 
that aim to ensure performance portability across a wide range of computer architectures.

For multicore processors, the conventional approach to exploit
parallelism in the dense matrix factorization (DMF) routines implemented
in LAPACK has relied, for many years, on the use of a multi-threaded BLAS (MTB). The list of 
current high performance instances of this
library composed of basic building blocks includes
Intel MKL~\cite{MKL}, IBM ESSL~\cite{ESSL}, 
GotoBLAS~\cite{Goto:2008:AHM:1356052.1356053,Goto:2008:HPI}, OpenBLAS~\cite{OpenBLAS}, ATLAS~\cite{atlas} or BLIS~\cite{BLIS1}.
These implementations exert a strict control over the data movements and can be expected to make an extremely efficient use of the cache memories.
Unfortunately, for complex DLA operations, 
this approach constrains the concurrency that can be leveraged by imposing an artificial fork-join model of execution
on the algorithm. Specifically, with
this solution, parallelism does not expand across multiple invocations to BLAS kernels even if they are independent
and, therefore, could be executed in parallel.

The increase in hardware concurrency of multicore processors in recent years 
has led to the development of parallel versions of
some DLA operations that exploit {\em task-parallelism} via a runtime (RTM).
Several relevant examples comprise the efforts with OmpSs~\cite{ompssweb}, PLASMA-Quark~\cite{plasmaweb}, StarPU~\cite{starpuweb}, Chameleon~\cite{chamaleonweb} and {\tt libflame}-SuperMatrix~\cite{flameweb}.
In short detail, the task-parallel RTM-assisted parallelizations decompose a DLA operation into a collection
of fine-grained tasks, interconnected with dependencies,
and issues the execution of each task to a single core,
simultaneously executing independent tasks on different cores while fulfilling the dependency constraints.
The RTM-based solution is better equipped to
tackle the increasing number of cores of current and future
architectures, because it leverages the natural concurrency
that is present in the algorithm. However, with this type of solution, the cores compete for
the shared memory resources and may not amortize completely the overhead of invoking the BLAS to perform fine-grain 
tasks~\cite{catalan17}.

In this paper we demonstrate that, for complex DMFs, it is possible to leverage the advantages of 
both approaches, extracting coarse-grain task-parallelism via a static look-ahead strategy~\cite{Str98},
with the multi-threaded execution of certain highly-parallel BLAS with fine granularity. 
Our solution thus exhibits some relevant differences with respect to an approach
based solely on either MTB or RTM,
making the following contributions:
\begin{itemize}
\item From the point of view of abstraction, 
      we use of a high-level parallel application programming interface (API), such as OpenMP~\cite{openmp},
      to identify two parallel sections (per iteration of the DMF algorithm) that become coarse-grain tasks to be run in parallel.
\item Within some of these coarse tasks, we employ OpenMP as well to extract loop-parallelism while strictly controlling the
      data movements across the cache hierarchy, yielding two nested levels of parallelism.
\item In contrast with a RTM-based approach, we apply a static version of look-ahead~\cite{Str98} (instead of a dynamic one), 
      in order to remove the panel factorization from the critical path of the algorithm's execution.
      This is combined with a cache-aware parallelization of the trailing update where all threads efficiently share
      the memory resources.
\item We offer a high-level description of the DMF algorithms, yet with enough details about their parallelization
      to allow the practical development of a library for dense linear algebra on multicore processors. 
\item We expose the distinct behaviors of the DMF algorithms on top of GNU's or Intel's OpenMP runtimes when dealing
      with nested parallelism on multicore processors. 
      For the latter, we illustrate how to correctly set a few environment variables that are
      key to avoid oversubscription and obtain high performance for DMFs.
\item We investigate the performance of the DMF algorithms when running on top
      an alternative multi-threading runtime based on the light-weight thread (LWT) library in Argobots~\cite{argobots}
      accessed via the OpenMP-compatible APIs GLT+GLTO~\cite{GLTAPI,lwthlpm}.
\item We provide a complete 
      experimental evaluation that shows the performance advantages of our approach using three representative
      DMF on a 8-core server with recent Intel Xeon technology.
\end{itemize}

The rest of the paper is organized as follows. 
In Section~\ref{sec:blas},
we review the cache-aware implementation and multi-threaded parallelization of the BLAS-3 in the BLIS framework.
In Section~\ref{sec:dmf}, we present a general framework that accommodates a variety of DMFs, elaborating on their
conventional MTB-based and the more recent RTM-assisted parallelization.
In Section~\ref{sec:OpenMP}, we present our alternative that combines task-loop parallelization,
static look-ahead, and a ``malleable'' instance of BLAS.
In Section~\ref{sec:LWT}, we discuss nested parallelism and
inspect the parallelization of DMF via the LWT runtime library underlying Argobots 
and the OpenMP APIs GLT and GLTO~\cite{argobots,GLTAPI,GLTO}.
Finally, in 
Section~\ref{sec:experiments} we provide an experimental evaluation of the different algorithms/implementations for
three representative DFMs, 
and in 
Section~\ref{sec:remarks} we close the paper with a few concluding remarks.


%% file: s2-gemm.tex
\section{Multi-threaded BLIS}
\label{sec:blas}

BLIS 
is a framework 
to develop high-performance implementations of BLAS and
BLAS-like operations on current architectures~\cite{BLIS1}.
We next review the design principles that underlie BLIS. For this purpose, we use the implementation of the 
general matrix-matrix multiplication (\gemm) in this framework/library in order to expose
how to exploit fine-grain 
{\em loop-parallelism} within the BLIS kernels, while carefully taking into account the cache organization.

\subsection{Exploiting the cache hierarchy}

Consider three matrices $A$, $B$ and $C$, of dimensions 
$m\times k$,
$k\times n$ and
$m\times n$, respectively.
BLIS mimics GotoBLAS 
to implement the \gemm operation%
\begin{equation}
\label{eqn:gemm}
C \mathrel{+}= A \cdot B
\end{equation}
(as well as variants of this operation with transposed/conjugate $A$ and/or $B$)
as three nested loops around a macro-kernel plus two packing routines;
see Loops~1--3 in Listing~\ref{lst:gotoblas_gemm}.
The macro-kernel is realized as two additional loops around a {\em
micro-kernel}; see Loops~4 and~5 in that listing.
In the code, 
$C_c (i_r:i_r+m_r-1,j_r:j_r+n_r-1)$
is a notation artifact, introduced to ease the presentation of the algorithm and no data copies are involved. 
In contrast, $A_c,B_c$ correspond to actual buffers that are involved in data copies.

The loop ordering in BLIS, together with the packing
routines and an appropriate choice of the cache configuration parameters $n_c$, $k_c$, $m_c$, $n_r$ and $m_r$,
dictate a regular movement of the data across the memory
hierarchy. Furthermore, these selections aim to
amortize the cost of these transfers with enough computation from within the micro-kernel to deliver high performance~\cite{BLIS1}.
In particular, BLIS is designed 
to maintain $B_c$ into the L3 cache (if present), $A_c$ into the L2 cache, and a micro-panel of $B_c$ (of dimension $k_c\times n_r$)  
into the L1 cache; in contrast,
$C$ is directly streamed from main memory to the core registers.
\begin{lstlisting}[float=th,language=C,caption=High performance implementation of \gemm in BLIS.,label=lst:gotoblas_gemm] 
void Gemm(int m, int n, int k, double *A, double *B, double *C) {
  // Declarations: mc, nc, kc,...
  for ( jc = 0; jc < n; jc += nc )                       // Loop 1
    for ( pc = 0; pc < k; pc += kc ) {                   // Loop 2
      // %*$B(p_c:p_c+k_c-1,j_c:j_c+n_c-1) \rightarrow \textcolor{blue}{B_c}$*)
      Pack_buffer_B(kc, nc, &B(pc,jc), &Bc);    
      for ( ic = 0; ic < m; ic += mc ) {                 // Loop 3
        // %*$A(i_c:i_c+m_c-1,p_c:p_c+k_c-1) \rightarrow \textcolor{red}{A_c}$*)
        Pack_buffer_A(mc, kc, &A(ic,pc), &Ac);  
        // Macro-kernel:
        for ( jr = 0; jr < nc; jr += nr )                // Loop 4
          for ( ir = 0; ir < mc; ir += mr ) {            // Loop 5
            // Micro-kernel: 
            //   %* $C_c(i_r:i_r+m_r-1,j_r:j_r+n_r-1) ~+=$ *)                 
            //   %* ~~~~$\textcolor{red}{Ac(i_r:i_r+m_r-1,1:1+k_c-1)} ~\cdot$*)                 
            //   %* ~~~~$\textcolor{blue}{Bc(j,1:1+k_c-1,_r:j_r+n_r-1)}$*)                 
            Gemm_mkernel( mr, nr, kc, &Ac(ir,1), &Bc(1,jr), 
                                                 &Cc(ir,jr) ); 
          }
      }
    }
}
\end{lstlisting}

\subsection{Multi-threaded parallelization}
\label{subsec:mtb}

The parallelization strategy of BLIS 
for multi-threaded architectures 
takes advantage of the loop-parallelism exposed by
the five nested-loop organization of \gemm at one or more levels.
A convenient option in most single-socket systems is to parallelize either Loop~3 (indexed by $i_c$), Loop~4 (indexed by $j_r$), or a combination of both~\cite{BLIS2,BLIS3,Catalan2016}.

For example, we can leverage the OpenMP parallel application programming interface (API)
to parallelize Loop~4 inside \gemm, with $t_{\textsc{mm}}$ threads, by inserting a simple
{\tt parallel for} directive before that loop (hereafter, for brevity, 
we omit most of the parts of the codes that do not experience any change with respect to their baseline reference):\\
\begin{minipage}[t]{\textwidth}
\begin{lstlisting}
// Fragment of Gemm: Reference code in Listing 1
void Gemm(int m, int n, int k, double *A, double *B, double *C) {
  // Declarations: mc, nc, kc,...
  for ( jc = 0; jc < n; jc += nc )                       // Loop 1
    // Loops 2, 3, 4 and packing of Bc, Ac (omitted for simplicity)
    // ...
        %*\color{red}{\#pragma omp parallel for num\_threads(tMM)*)
        for ( jr = 0; jr < nc; jr += nr )                // Loop 4
          // Loop 5 and GEMM micro-kernel (omitted)
          // ...
}
\end{lstlisting}
\end{minipage}

\begin{figure*}[tb!]
\centering %
\includegraphics[width=\textwidth]{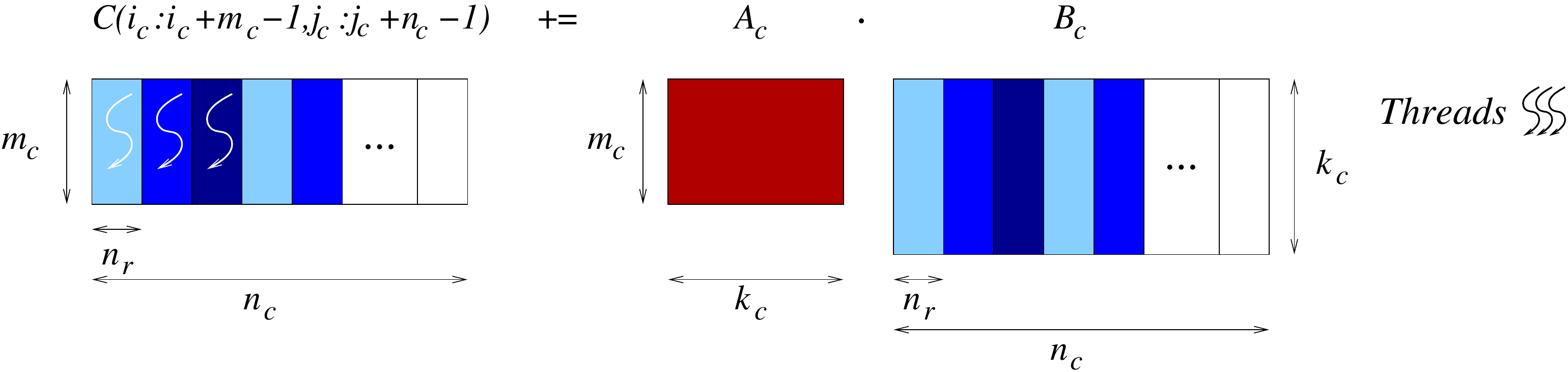}
\caption{Distribution of the workload among $t_{\textsc{mm}}=3$~threads when Loop~4 of BLIS \gemm is parallelized. Different colors in the output
$C$ distinguish the micro-panels of this matrix that are computed by each thread as the product of $A_c$ and corresponding micro-panels of the input $B_c$.}
  \label{fig:parallel_loop4}
\end{figure*}

Unless otherwise stated, in the remainder of the paper we will consider a version of BLIS \gemm that extracts loop-parallelism from Loop~4 only,
using $t_{\textsc{mm}}$ threads;
see Figure~\ref{fig:parallel_loop4}.
To improve performance, the packing of $A_c$ and $B_c$ are also performed in parallel so that, for example,
at each iteration of Loop~3, all $t_{\textsc{mm}}$~threads collaborate to copy and re-organize
the entries of $A(i_c:i_c+m_c-1,p_c:p_c+k_c-1)$ into the buffer $A_c$.
From the point of view of the cache utilization, with this parallelization strategy, 
all threads share the same buffers $A_c$ and $B_c$, while each thread operates on a distinct
micro-panel of $B_c$, of dimension $k_c \times n_r$.
The shared buffers for $A_c, B_c$ are stored in the L2, L3 caches while the micro-panels of $B_c$ reside in the L1
cache. 

%% file: s3-dmf.tex
\section{Parallel Dense Matrix Factorizations}
\label{sec:dmf}

\input{MF_blk}

\subsection{A general framework}

Many of the routines for DMFs in LAPACK
fit into a common algorithmic skeleton, consisting of a loop that processes
the input matrix in steps of $\bs$ columns/rows per iteration. 
In general the parameter $\bs$ is referred to as the algorithmic block size.
We next offer a general framework that accommodates the routines for the LU, Cholesky, QR and LDL$^T$ 
factorizations (as well as matrix inversion via Gauss-Jordan elimination)~\cite{GVL3}. To some extent, it also applies 
to two-sided decompositions for the reduction to compact band forms in two-stage methods
for  the solution of eigenvalue problems and 
the computation of the singular value decomposition (SVD)~\cite{Bischof:2000:AST}.

Let us
denote the input $m \times n$ matrix to factorize
as $A$,
and assume, for simplicity,
that $m=n$ and this dimension is an integer multiple of the block size $\bs$.
Many routines for the afore-mentioned DMFs (and matrix inversion) fit into the general code skeleton 
displayed in Listing~\ref{lst:dmf_flame_algorithm}, which is 
partially based on the FLAME API for the C programming language~\cite{FLAME:Recipe}.  
In that scheme, before the loop commences, and in preparation for the first iteration,
routine {\tt FLA\_Part\_2x2} decouples the input matrix
as\\
\begin{center}
\partitionings
\quad where
\partitionsizes.
\end{center}
~\\
This initial partition thus enforces that
$A\equiv A_{BR}$ while the remaining three blocks ($A_{TR}, A_{BL}, A_{BR}$) are void.

Inside the loop body, at the beginning of each iteration, 
routine {\tt FLA\_Repart\_2x2\_to\_3x3} 
performs a new decoupling: 
\begin{center}
\repartitionings
\quad where
\repartitionsizes.
\end{center}
~\\
This partition exposes
the {\em panel} (column block)
$\left(
\begin{array}{c}
A_{11} \\\hline
A_{21}
\end{array}
\right),$
consisting of $\bs$ columns,
and the
{\em trailing submatrix}
$\left(
\begin{array}{c}
A_{12} \\\hline
A_{22}
\end{array}
\right).$

After the {\tt Operations}, the loop body is closed 
by routine {\tt FLA\_Cont\_with\_3x3\_to\_2x2},
which realizes a repartition artifact\\
\begin{center}
\moveboundaries,
\end{center}
~\\
advancing the boundaries (thick lines) within the matrix by $b$ rows/columns,
in preparation for the next iteration.


\begin{lstlisting}[float=th,language=C,caption=Skeleton routine for a DMF.,label=lst:dmf_flame_algorithm]
void FLA_DMF( int n, FLA_Obj A, int b )
{
  // Declarations: ATL, ATR,..., A00, A01,... are FLA_Obj(ects)

  // Partition matrix into 2 x 2, with ATL of dimension 0 x 0
  FLA_Part_2x2( A,    &ATL, &ATR,
                      &ABL, &ABR,     0, 0, FLA_TL );

  for ( k = 0; k < n / b; k++ ) {

    // Repartition 2x2 -> 3x3 with A11 of dimension b x b
    FLA_Repart_2x2_to_3x3(
       ATL, /**/ ATR,       &A00, /**/ &A01, &A02,
    /* ************* */  /* ********************* */
                            &A10, /**/ &A11, &A12,
       ABL, /**/ ABR,       &A20, /**/ &A21, &A22,
       b, b, FLA_BR );
    /*-----------------------------------------------------------*/
    // Operations
    // ...
    /*-----------------------------------------------------------*/
    // Move boundaries 2x2 <- 3x3 in preparation for next iteration
    FLA_Cont_with_3x3_to_2x2(
       &ATL, /**/ &ATR,        A00, A01, /**/ A02,
                               A10, A11, /**/ A12,
    /* *************** */   /* ****************** */
       &ABL, /**/ &ABR,        A20, A21, /**/ A22,
       FLA_TL );
  }
}
\end{lstlisting}

%

In the blocked right-looking variants of the DMF routines, inside the loop body for the iteration, the 
current panel is factorized 
and the transformations employed for this purpose are applied to the trailing submatrix:\\
\begin{minipage}[t]{\textwidth}
\begin{lstlisting}[language=C]
    // Fragment of FLA_DMF: Reference code in Listing 2
    /*-----------------------------------------------------------*/
    // Operations
        PF( &A11,            // Panel factorization
            &A12 );          
        TU( &A11, &A21,      // Trailing update
            &A12, &A22 );  
    /*-----------------------------------------------------------*/
\end{lstlisting}
\end{minipage}

For high performance, the width of the panel (i.e., its number of columns, $\bs$) 
is in general set to a small value (a few hundreds) 
in order to cast most computations in terms of the compute-intensive trailing update. 

To conclude the presentation of the general framework for DMF, hereafter we will abstract many of the details
in the DMF routine, to obtain a simpler algorithm as that shown in Listing~\ref{lst:dmf_algorithm}.
For simplicity, we omit there the partitioning operations and replace the operands passed to the panel factorization and trailing update by the iteration index {\tt k}.

\begin{lstlisting}[float=th,language=C,caption=Simplified routine for a DMF.,label=lst:dmf_algorithm]
void FLA_DMF( int n, FLA_Obj A, int b )
{
  for ( k = 0; k < n / b; k++ ) {
    /*-----------------------------------------------------------*/
    // Operations
        PF( k );             // Panel factorization
        TU( k );             // Trailing update 
    /*-----------------------------------------------------------*/
  }
}
\end{lstlisting}

\subsection{Exploiting loop-parallelism via MTB}

For high performance, the DMF routines in LAPACK cast most of their computations in terms of the BLAS.
Therefore, for many years,
the conventional approach to extract parallelism from these routines has simply linked 
them with  a multi-threaded instance of the latter library;
see Section~\ref{sec:blas}.
For the DMFs, the panel factorization is generally decomposed into fine-grain kernels, 
some of them realized via calls to the BLAS. 
The same occurs for the trailing update though, in this case, this operation involves 
larger matrix blocks and rather simple dependencies. In consequence there is a considerable greater amount 
of concurrency in the trailing update compared with that present in the panel factorization.
For a few decades, the MTB approach has reported reasonable performance for DMFs, at a minimal tuning effort,
provided a highly-tuned implementation of the BLAS was available for the target architecture.

\subsection{Exploiting task-parallelism via RTM}

The RTM approach exposes task-parallelism by decomposing the trailing update into multiple tasks, controlling the dependencies
among these tasks, and simultaneously executing independent tasks in different cores. This is illustrated in 
Listing~\ref{lst:dmf_task_parallel_algorithm}, using the OpenMP parallel programming API. Note how 
the {\tt k}-th trailing update operation
\tu$_k$ is divided there into multiple panel updates, 
$\tu_k \rightarrow ( \tu_k^{k+1}~\vert~\tu_k^{k+2}~\vert~\tu_k^{k+3}\ldots )$.
These tasks are then processed inside the loop indexed by variable {\tt j} via successive calls to routine {\tt TU\_panel}.
For clarity, the parallelization exposed 
in the code contains a simplified mechanism for the detection of dependencies, which should be 
specified in terms of the actual operands
instead of their indices.
In short detail, a dependency with respect to panel {\tt j} can be, e.g., specified in terms of the
top-left entry of the {\tt j}-th panel, which can act as a ``representant'' for all the elements in 
that block~\cite{BadiaHLPQQ09}.

\begin{lstlisting}[float=th,language=C,caption=Task-parallel routine for a DMF using OpenMP.,label=lst:dmf_task_parallel_algorithm]
void FLA_DMF_task_parallel( int n, FLA_Obj A, int b )
{
%*\color{red}{\#pragma omp parallel}*)
%*\color{red}{\#pragma omp single}*)
%*\color{red}{\{*)
  for ( k = 0; k < n / b; k++ ) {
    /*-----------------------------------------------------------*/
    // Operations
        %*\color{red}{\#pragma omp task depend( inout:k )}*)
        PF( k );             // Panel factorization
        for ( j = k+1; j < n / b; j++ ) {
          %*\color{red}{\#pragma omp task depend( in:k ) depend( inout:j )}*)
          TU_panel( k, j );    // Trailing update of panel
        }
    /*-----------------------------------------------------------*/
  }
%*\color{red}{\}*)
}
\end{lstlisting}

For several DMFs, the RTM can also decompose the panel factorization into
multiple tasks, in an attempt to remove this operation from the critical path of the 
algorithm~\cite{Buttari200938,Quintana-Orti:2009:PMA:1527286.1527288}.
However, for some DLA operations such as the LU factorization with partial pivoting (LUpp), 
performing that type of task decomposition 
requires a different pivoting strategy, which
modifies the numerical properties of the algorithm~\cite{Quintana-Orti:2008}.

\subsection{Performance of MTB vs RTM}
\label{subsec:MTBvsRTM}

We next expose the practical performance of the MTB and RTM parallelization approaches 
using two representative DLA operations: \gemm and LUpp.
For these experiments we employ 
an 8-core Intel Xeon E5-2630 v3 processor, 
Intel's {\tt icc} runtime, and BLIS 0.1.8 with 
the cache configuration parameters set to 
optimal values for the Intel Haswell architecture.
(The complete details about the experimentation setup are given in Section~\ref{sec:experiments}.)

Our MTB version of \gemm (MTB-\gemm) simply extracts parallelism from Loop~4 and the packing routines, 
as described in subsection~\ref{subsec:mtb}.
Assuming all three matrix 
operands for the multiplication are square of dimension $n$, and this value is an integer multiple of $\bs$,
the task-parallel RTM code (RTM-\gemm) divides the three matrices into square $\bs \times \bs$ blocks, 
so that 
\[
 C_{ij} = \sum_{k=0}^{n/\bs-1}
 A_{ik} \cdot
 B_{kj}, \quad i,j = 0,1,\ldots,n/\bs-1,
\]
and specifies each one of the
smaller operations
$C_{ij} +=A_{ik}\cdot B_{kj}$ as a task.

The MTB version of LUpp (MTB-LU) corresponds to the reference routine
{\sc getrf} in the implementation of LAPACK in netlib.\footnote{http://www.netlib.org/lapack}
At each iteration, the code first computes the panel factorization ({\sc getf2}) to next update the trailing submatrix via a row permutation 
({\sc laswp}), followed by a triangular system solve
({\sc trsm}) and a matrix-matrix multiplication
(\gemm).
Parallelism is extracted via the multi-threaded versions of the latter two kernels in BLIS and  
a simple column-oriented multi-threaded implementation 
of the row permutation routine parallelized using OpenMP. 
The RTM version of LUpp (RTM-LU) specifies the panel factorization arising at each iteration as a task, and
``taskifies'' the trailing update into column panels, as described in the generic code in Listing~\ref{lst:dmf_task_parallel_algorithm}.
The blocking parameter is set to $\bs$=192
as this value matches the optimal 
$k_c$ for the target architecture and, therefore, can be expected to enhance the performance of the micro-kernel~\cite{BLIS1}.



\begin{figure}[htb]
\centering
\includegraphics[width=0.6\textwidth]{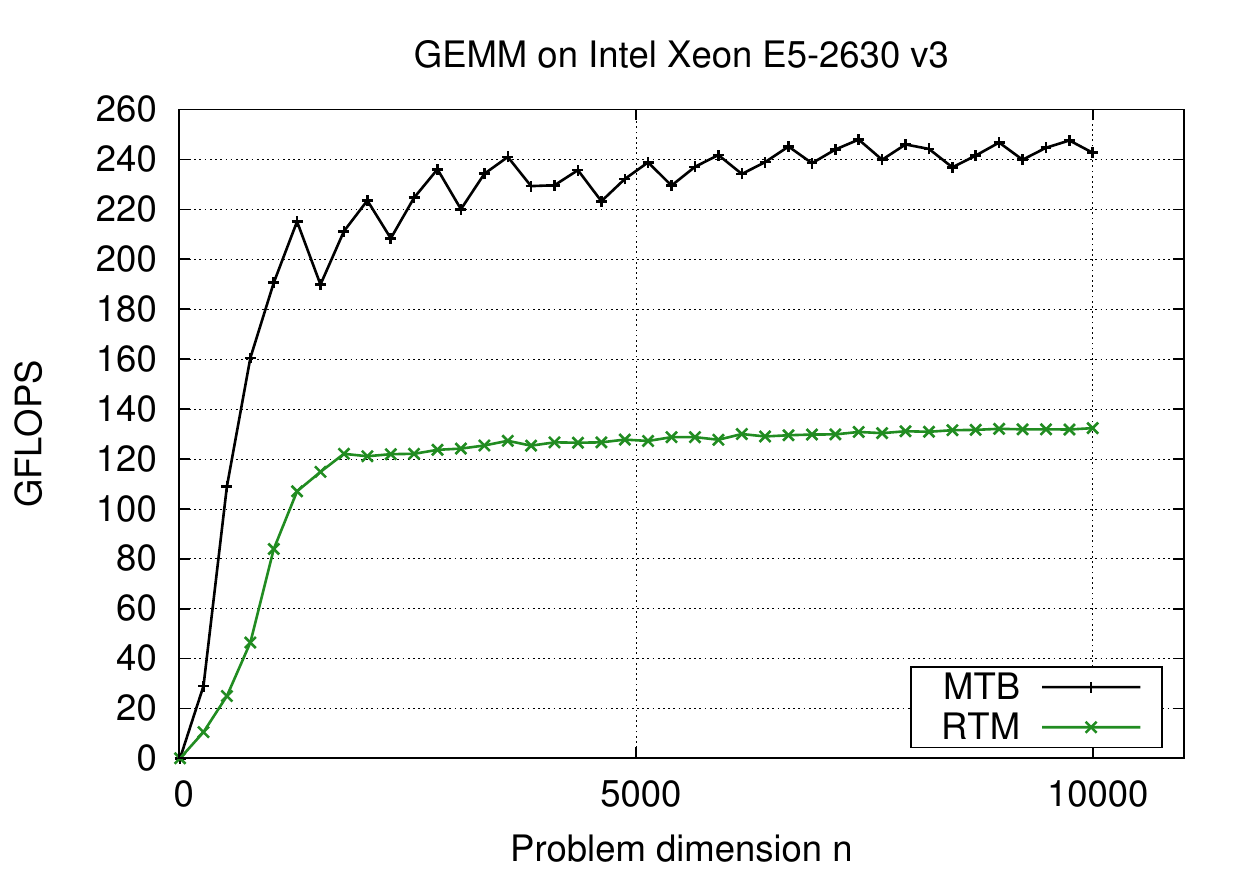}
\includegraphics[width=0.6\textwidth]{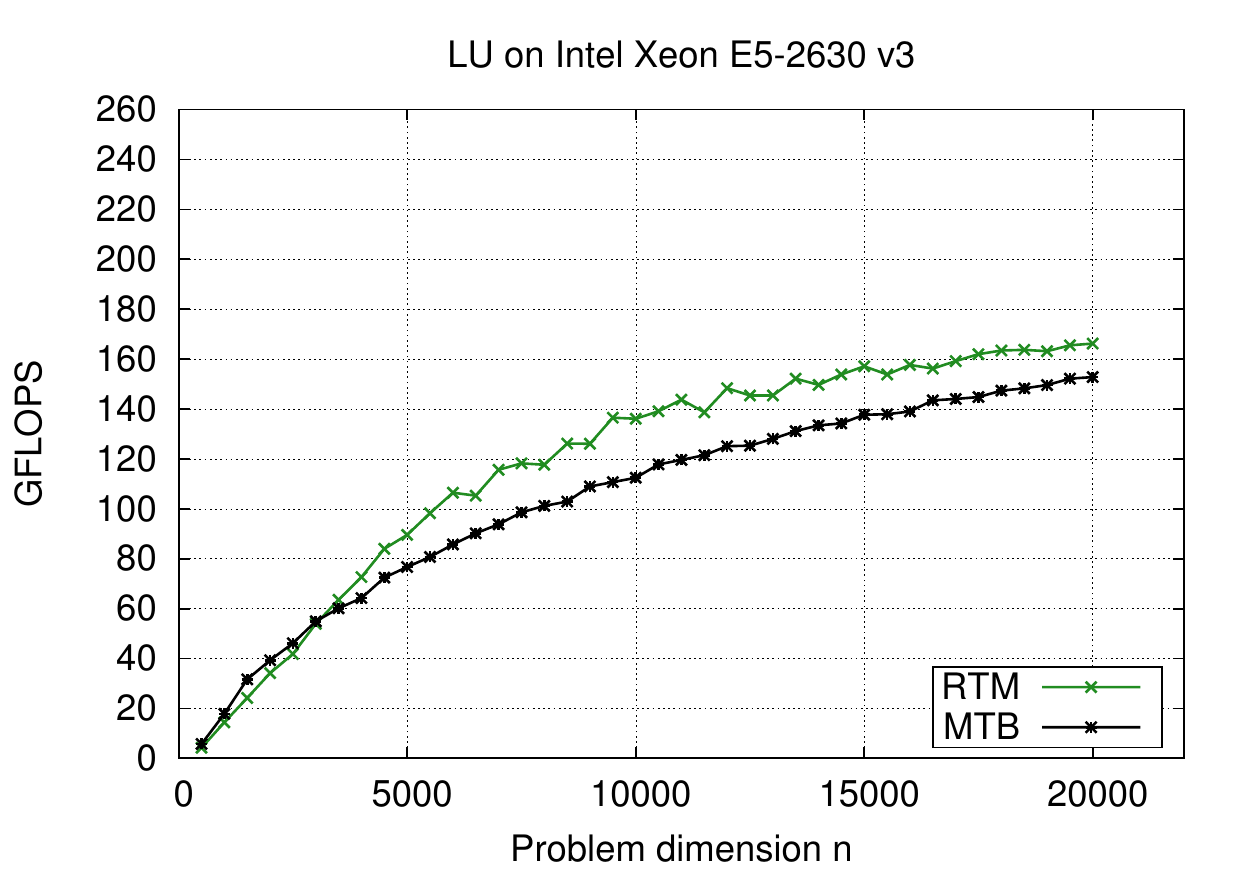}
\caption{Performance of \gemm (top) and LUpp (bottom) using MTB vs RTM.}
\label{fig:parallel_gemm_and_LU}
\end{figure}

Figure~\ref{fig:parallel_gemm_and_LU}
reports the GFLOPS (billions of flops per second) rates attained by the MTB and RTM parallelizations of \gemm and LUpp
using all 8~cores.  
The results in the top plot show that MTB-\gemm (which corresponds to a single call to the \gemm routine in BLIS) delivers 
up to 245~GFLOPS.
Compared with this, when we decompose this highly-parallel operation into multiple tasks, and use 
Intel's OpenMP RTM to exploit this type of parallelism, the result
is a considerable drop in the performance rate. 
The reason is that, for RTM-\gemm, the threads compete for the shared cache memory levels, and
the packing and the RTM overheads become more visible.

The LUpp factorization presents the opposite behavior. In this case, MTB-LU suffers
from the adoption of the  fork-join parallelization model, where the threads become active/blocked at the 
beginning/end of each invocation to BLAS. In consequence, parallelism cannot be exploited across distinct BLAS kernels and the panel factorization
becomes a performance bottleneck~\cite{catalan17}.
RTM-LU overcomes this problem by introducing a sort of dynamic look-ahead strategy that can overlap the execution of the ``future'' panel
factorization(s) with that of the ``current'' trailing update~\cite{Buttari200938,Quintana-Orti:2009:PMA:1527286.1527288}.
The result is a performance rate that, 
for large problems, is higher than that of MTB-LUpp but still far below that of MTB-\gemm, 
especially for small and moderate problem dimensions.


\subsection{Impact for DMFs}

\begin{figure}
\includegraphics[width=\textwidth]{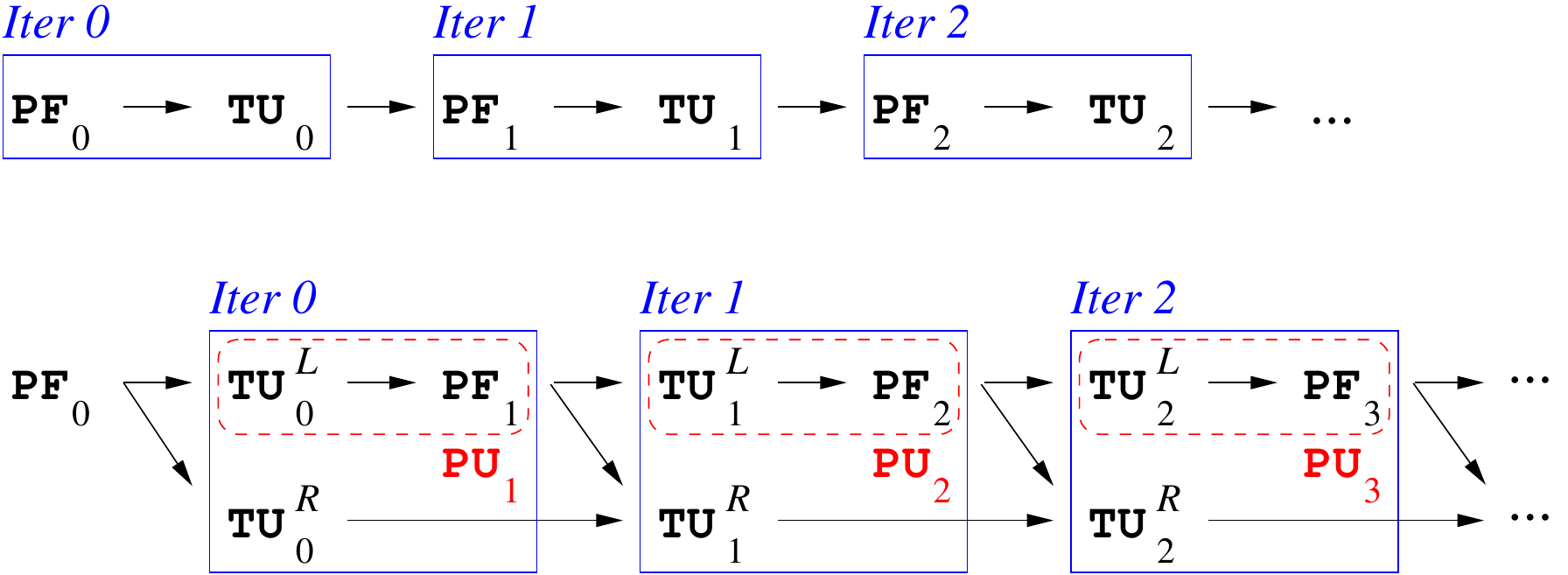}
\caption{Dependencies in the blocked right-looking algorithms for DMFs without and with look-ahead (top and bottom, respectively). Following the convention,
         $\pf$ stands for panel factorization and
         $\tu$ for trailing update; the subindices simply refer to the iteration index {\tt k}; see, e.g., Listing~\ref{lst:dmf_algorithm}.}
\label{fig:dependencies}
\end{figure}

Let us re-consider the dependencies appearing in the DMF algorithms.
The partitions of the general algorithm in Listing~\ref{lst:dmf_algorithm}, and the operations present in the blocked right-looking algorithm, 
determine a dependency
acyclic graph (DAG) with the structure illustrated in Figure~\ref{fig:dependencies} (top).
This DAG also exposes the problem represented by the panel factorization in MTB-LU (or any other DMF parallelized with the same strategy).
As the number of cores grows, the relative cost of the highly-parallel trailing update is reduced,
transforming the largely-sequential panel factorization into a major performance bottleneck. 
The RTM-LU parallelization attacks this problem by dividing the trailing update into multiple panels/suboperations (or tasks)
$\tu_k \rightarrow ( \tu_k^{k+1}~\vert~\tu_k^{k+2}~\vert~\tu_k^{k+3}\ldots )$ 
and overlapping their modification with that of future panel factorizations. 
In exploiting this task-parallelism, however, it breaks the
highly-parallel trailing update into multiple operations, to be computed by a collection of threads that compete for the shared 
memory resources.

The discussion in this section emphasizes two insights that we can summarize as follows:
\begin{itemize}
\item The trailing update is composed of highly-parallel and simple kernels from BLAS that could profit from a fine-grain control of the cache hierarchy for
high performance. 
\item The panel factorization, in contrast, is mostly sequential and needs to be overlapped with the trailing update to prevent 
it from becoming a 
bottleneck for the performance of the global algorithm.
\end{itemize}

%% file: MF_blk.tex
\resetsteps      


\renewcommand{\routinename}{ \left[ A \right] := \mbox{\sc LU\_blk}( A ) }


\renewcommand{\guard}{
  n( A_{TL} ) < n( A )
}


\renewcommand{\partitionings}{
  $
  A \rightarrow
  \FlaTwoByTwo{A_{TL}}{A_{TR}}
              {A_{BL}}{A_{BR}}
  $
}

\renewcommand{\partitionsizes}{
$ A_{TL} $ is $ 0 \times 0 $
}


\renewcommand{\blocksize}{b}

\renewcommand{\repartitionings}{
$  \FlaTwoByTwo{A_{TL}}{A_{TR}}
              {A_{BL}}{A_{BR}}
  \rightarrow
  \FlaThreeByThreeBR{A_{00}}{A_{01}}{A_{02}}
                    {A_{10}}{A_{11}}{A_{12}}
                    {A_{20}}{A_{21}}{A_{22}}
$}

\renewcommand{\repartitionsizes}{
  $ A_{11} $ is $ b \times b $}


\renewcommand{\moveboundaries}{
$  \FlaTwoByTwo{A_{TL}}{A_{TR}}
              {A_{BL}}{A_{BR}}
  \leftarrow
  \FlaThreeByThreeTL{A_{00}}{A_{01}}{A_{02}}
                    {A_{10}}{A_{11}}{A_{12}}
                    {A_{20}}{A_{21}}{A_{22}}
$}


\renewcommand{\update}{
$
  \begin{array}{cccl}
    {\sf RL1.} & \left[ \begin{array}{c}  A_{11}\\ A_{21} \end{array}  \right] &:=& 
         \mbox{\sc LU\_unb}\left( \left[ \begin{array}{c}  A_{11}\\ A_{21}  \end{array} \right] \right)\\ [0.15in]
    {\sf RL2.} & A_{12} &:=& 
                \trilu{A_{11}}^{-1} A_{12} \\
    {\sf RL3.} & A_{22} &:=& A_{22} - A_{21} A_{12}\\
  \end{array}
$
}


%% file: s4-openmp.tex
\section{Static Look-ahead and Mixed Parallelism}
\label{sec:OpenMP}



The introduction of static look-ahead~\cite{Str98} aims to overcome the strict dependencies 
in the DMF. 
For this purpose, the following modifications are introduced into the conventional factorization algorithm:
\begin{itemize}
\item The trailing update is broken into two panels/suboperations/tasks only,
$\tu_k \rightarrow ( \tu_k^L~\vert~\tu_k^R )$, where 
$\tu_k^L$ contains the leftmost $\bs$ columns of $\tu_k$, which exactly overlap with those of $\pf_{k+1}$.
\item The algorithm is then (manually) re-organized, applying a sort of software pipelining in order to perform 
the panel factorization $\pf_{k+1}$ in the same iteration 
as 
the update $( \tu_k^L~\vert~\tu_k^R )$. 
\end{itemize}
These changes allow to overlap the sequential factorization of the ``next'' panel
with the highly parallel update of the ``current'' trailing submatrix in the same iteration; see Figure~\ref{fig:dependencies} (bottom)
and the re-organized version of the DMF with look-ahead in Listing~\ref{lst:dmf_algorithm_la}.
There, we assume that 
the {\tt k}-th left trailing update $\tu_k^L$ 
and the $(k+1)$-th panel factorization $\pf_{k+1}$
are both performed inside
routine 
{\tt PU( k+1 )} (for panel update); 
and the {\tt k}-th right trailing update $\tu_k^R$ occurs inside
routine 
{\tt TU\_right( k )}.





\begin{lstlisting}[float=th,language=C,caption=Simplified routine for a DMF with look-ahead.,label=lst:dmf_algorithm_la]
void FLA_DMF_la( int n, FLA_Obj A, int b )
{
  PF( 0 );                  // First panel factorization
  for ( k = 0; k < n / b; k++ ) {
    /*-----------------------------------------------------------*/
    // Operations
        PU( k+1 );           // Panel update: PF + TU (left)
        TU_right( k );       // Trailing update (right)
    /*-----------------------------------------------------------*/
  }
}
\end{lstlisting}



\subsection{Parallelization with the OpenMP API}
\label{subsec:gemm_dmf}

The goal of our ``mixed'' strategy exposed next is to exploit a combination of task-level and loop-level parallelism in the static look-ahead variant, 
extracting coarse-grain task-level parallelism between the independent tasks 
$\pu_{k+1}$ and 
$\tu_{k}^R$
at each iteration, while
leveraging the fine-grain loop-parallelism within
the latter using a cache-aware multi-threaded implementation of the BLAS.


Let us assume that, for an architecture with $t$ hardware cores, we 
want to spawn one OpenMP thread per core, with a single thread
dedicated to the panel update 
$\pu_{k+1}$ and the remaining $t_{\textsc{mm}} = t-1$ to the right trailing update
$\tu_{k}^R$.
(This mapping of tasks to threads aims to match
the reduced and ample degrees of parallelism of the panel factorization (inside the panel update) and trailing update, respectively.)
To attain this objective, we can then use the OpenMP {\tt parallel sections} directive
to parallelize the operations in the loop body of the algorithm for the DMF as follows:\\
%
\begin{minipage}[t]{\textwidth}
\begin{lstlisting}
    // Fragment of FLA_DMF_la: Reference code in Listing 5
    /*-----------------------------------------------------------*/
    // Operations
    %*\color{red}{tMM = t-1;}*)
    %*\color{red}{\#pragma omp parallel sections num\_threads(2)}*)
    %*\color{red}{\{*)
      %*\color{red}{\#pragma omp section*)
        PU( k+1 );           // Panel update: PF + TU (left)
      %*\color{red}{\#pragma omp section*)
        TU_right( k );       // Trailing update (right)
    %*\color{red}{\}*)
    /*-----------------------------------------------------------*/
\end{lstlisting}
\end{minipage}
~\\
Here we map the panel update and trailing update to one thread each. 
Then, the invocation to a loop-parallel instance of the BLAS from the trailing update (but a sequential one for the panel update)
yields the desired {\em nested-mixed parallelism} (NMP), with
the OpenMP {\tt parallel sections} directive at the ``outer'' level and a loop-parallelization of the BLAS (invoked from the right trailing update)
using OpenMP {\tt parallel for} directives at the ``inner'' level; see subsection~\ref{subsec:mtb}.

\subsection{Workload balancing via malleable BLAS}

Extracting parallelism within the iterations via a static look-ahead using the OpenMP {\tt parallel sections} directive implicitly sets a synchronization
point at the end of each iteration. In consequence, a performance bottleneck may appear if the practical costs 
(i.e., execution time) of $\pu_{k+1}$(=$\tu_{k}^R+\pf_k$) and $\tu_{k}^R$ are unbalanced.

A higher cost of $\pu_{k+1}$ 
is, in principle, due to the use of a value for $\bs$ that is too large and occurs when the number of cores
is relatively large with respect to the problem dimension. This can be alleviated by 
adjusting, on-the-fly, the block dimension via an auto-tuning technique referred to as {\em early termination}~\cite{catalan17}.

Here we focus on the more challenging opposite case, in which $\tu_{k}^R$ is the most expensive operation.
This scenario is tackled in~\cite{catalan17} by developing a {\em malleable thread-level} (MTL) implementation of the BLAS so that,
when the thread in charge of $\pu_{k+1}$ completes this task, it joins the remaining $t_{\textsc{mm}}$ threads that are executing
$\tu_{k}^R$. Note that this is only possible because the instance of BLAS that we are using is open source, and in consequence, we can
modify the code to achieve the desired behavior.
In comparison, standard multi-threaded instances of BLAS, such as those in Intel MKL, OpenBLAS or GotoBLAS, allow the user to run
a BLAS kernel with a certain amount of threads, but this number cannot be varied during the execution of the kernel (that is on-the-fly).

Coming back to our OpenMP-based solution, we can attain the malleability effect as follows:\\
\begin{minipage}[t]{\textwidth}
\begin{lstlisting}
    // Fragment of FLA_DMF_la: Reference code in Listing 5
    /*-----------------------------------------------------------*/
    // Operations
    %*\color{red}{tMM = t-1;}*)
    %*\color{red}{\#pragma omp parallel sections num\_threads(2)}*)
    %*\color{red}{\{*)
      %*\color{red}{\#pragma omp section*)
      %*\color{red}{\{*)
        PU( k+1 );           // Panel update: PF + TU (left)
        %*\color{red}{tMM = t;}*)
      %*\color{red}{\}*)
      %*\color{red}{\#pragma omp section*)
        TU_right( k );       // Trailing update (calls GEMM)
    %*\color{red}{\}*)
    /*-----------------------------------------------------------*/
\end{lstlisting}
\end{minipage}

For simplicity, let us assume the right trailing update boils down to a single call to \gemm.
Setting variable {\tt tMM=t} 
after the completion of the panel update 
(in line~8) 
ensures that,
provided this change is visible inside \gemm, the next time the 
OpenMP {\tt parallel for} directive around Loop~4 in \gemm is encountered 
(i.e., in the next iteration of Loop~3; see Listing~\ref{lst:gotoblas_gemm}), this loop will be executed by all 
$t$ threads.  The change in the number of threads also affects the parallelism degree of the packing routine for $A_c$.




%% file: s5-lwt.tex
\section{Re-visiting Nested Mixed Parallelism}
\label{sec:LWT}

Exploiting data locality is crucial on current architectures. This is the case for many scientific applications and, especially, for DMF when the goal is to squeeze the last
drops of performance of an algorithm--architecture pair. 
To attain this, a tight control of the data placement/movement and threading activity may be necessary. Unfortunately, 
the use of a high-level programming model such as OpenMP abstracts these mappings, 
making this task more difficult.

\subsection{Conventional OS threads}
\label{subsec:nested}

Nested parallelism may potentially yield a performance issue due to the thread management 
realized by the underlying OpenMP runtime. 
In particular, when the first {\tt parallel} directive is found, a team of threads is created and the following region 
is executed in parallel.  Now, if a second {\tt parallel} directive is encountered inside the region (nested parallelism), 
a new team of threads is created for each thread encountering it. 
This runtime policy may spawn more threads than physical cores, adding a relevant overhead due to oversubscription
as current OpenMP releases are implemented on top of ``heavy'' Pthreads, which are controlled by the operating system (OS).   

In the DMF algorithms, we encounter nested parallelism because of the nested invocation of a
{\tt parallel for} (from a BLAS kernel) inside a {\tt parallel sections} 
directive (encountered in the DMF routine). To tackle this problem, we can restrict the number of threads for the {\tt sections} to only two and, 
in an architecture with {\tt t} physical cores, set the number of threads in 
the {\tt parallel for} to {\tt tMM}={\tt t}$-1$, for a total of {\tt t} threads. Unfortunately, 
with the addition of malleability, the thread that executes the panel factorization, upon completing this
computation, will remain ``alive'' (either in a busy wait or blocked) while a new thread is spawned for the next iteration of 
Loop~3 in the panel update, yielding a total of~{\tt t}$+1$ threads and the undesired oversubscription problem.

We will explore the practical effects of oversubscription for classical OpenMP runtimes that leverage OS threads
in Section~\ref{sec:experiments}, 
where we consider the differences between the OpenMP runtimes underlying
GNU {\tt gcc} and Intel {\tt icc} compilers, and describe how to avoid the negative consequences for the latter.


\subsection{LWT in Argobots}

In the remainder of this section we introduce an alternative to deal with oversubscription problems using 
the implementation of LWTs in Argobots~\cite{argobots}.
Compared with OS threads, LWTs (also known as user-level threads or ULTs) run in the user space,
providing a lower-cost threading mechanism (in terms of context-switch, suspend, cancel, etc.) than Pthreads~\cite{stein1992}.
Furthermore, LWT instances follow a two-level hierarchical implementation, where the bottom level (closer to the hardware) 
comprises the OS threads which are bound to cores following a 1:1 relationship. In contrast, the top level 
corresponds to the ULTs,
which contain the concurrent code that will be executed concurrently by the OS threads. With this strategy,
the number of OS threads will never exceed the amount of cores and, therefore, oversubscription is prevented. 





\subsubsection{LWT parallelization with GLTO}

To improve code portability, we utilize the GLTO API~\cite{GLTO}, which is an OpenMP-compatible implementation built on top of the 
GLT API~\cite{GLTAPI}, and rely on Argobots as the underlying threading library.
Concretely, our first LTW-based parallelization employs GLTO to extract task-parallelism from the DMF,
using the OpenMP {\tt parallel sections} directive,
and loop-parallelism inside the BLAS, using the OpenMP {\tt parallel for} directive.
Therefore, no changes are required to the code for the DMF with static look-ahead, NMP and MTL BLAS. 
The only difference is that the OpenMP threading library is replaced by GLTO's (i.e., Argobot's) instance in 
order to avoid potential oversubscription problems. 

Applied to the DMFs, this solution initially spawns one OS thread per core. The master thread first encounters the 
{\tt parallel sections} directive, creating two ULT work-units (one per section), 
and then commences the execution of one of these sections/ULTs/branches. 
Until the creation of the additional ULTs, the remaining threads cycle in a busy-wait. 
Once this occurs, one of these threads will commence with the execution of the alternative section (while the remaining ones will remain
in the busy-wait).
The thread in charge of the right trailing update then creates several ULTs inside the BLAS, one per iteration chunk due to
the {\tt parallel for} directive. These ULTs will be executed, when ready, by the OS threads.
The TLM technique is easily integrated in this solution as OS threads execute ULTs, independently of which section of the code they ``belong to''.
\begin{lstlisting}[float=th,language=C,label=lst:lllwt_gemm, caption=High performance implementation of \gemm in BLIS on top of GLT using Tasklets.]
void Gemm_Tasklets(int m, int n, int k, double *A, double *B, 
                                        double *C) {
  // Declarations: mc, nc, kc,...
  // GLT tasklet handlers
  %*\color{red}{GLT\_tasklet tasklet[tMM];}*)
  %*\color{black}{struct L4\_args L4args[tMM];}*)

  for ( jc = 0; jc < n; jc += nc ) {                     // Loop 1
    // Loops 2, 3, 4 and packing of Bc, Ac (omitted for simplicity)
        for ( th = 0; th < tMM; th++ )                  // Loop 4
        {
            %*\color{black}{L4args[th].arg1 = arg1;}*)
            %*\color{black}{L4args[th].arg2 = arg2;}*)
            // ...
            // Tasklet creation that invokes L4 function
            %*\color{red}{glt\_tasklet\_create(L4, L4args[th], \&tasklet[th]); *)
        }
    
        %*\color{red}{glt\_yield(); *)
        // Join the tasklets
        for ( th = 0; th < tMM; th++ )
              %*\color{red}{glt\_tasklet\_join(\&tasklet[th]); *)
  }
}
\end{lstlisting}

%
%
%
%

\subsubsection{LWT parallelization with GLTO+GLT}

Argobots provides direct access to Tasklets, a type of work-units that is even lighter than ULTs
and can deliver higher performance for just-computation codes~\cite{cluster16}. 
In our particular example, Tasklets can
leveraged to parallelize the BLAS routines, providing an MTL  black-box implementation of this library
that can be invoked from higher-level operations, such as DMFs.
In this alternative LWT-based parallel solution, 
the potential higher performance derived from the use of Tasklets comes at the cost of some development effort.
The reason is that GLTO does not support Tasklets but relies on ULTs to realize all work-units. Therefore, our implementation of
MTL BLAS has to abandon GLTO, employing the GLT API to introduce the use of Tasklets in the BLAS instance. 

In more detail, we implemented a hybrid solution with GLTO and GLT. At the outer level, the parallelization of the DMF employs 
the {\tt parallel sections} directive on top of GLTO, the OpenMP runtime and Argobots' threading mechanism. 
Internally, the BLAS routines are implemented with {\tt GLT\_tasklets},
as depicted in the example in Listing~\ref{lst:lllwt_gemm}. 
In the {\tt Gemm\_Tasklets} routine there, in line~5 we first declare the tasklet handlers (one per thread that will execute Loop~4, that
is, {\tt tMM}).
The original Loop~4 in {\tt Gemm}, indexed by {\tt jr} (see Listing~\ref{lst:gotoblas_gemm}), is then replaced by a loop that creates one
Tasklet per thread. Lines 12--14 inside this new loop initialize the arguments to function {\tt L4}, among other parameters
defining 
which iterations of the iteration space of the original
loop indexed by {\tt jr} will be executed as part of the Tasklet indexed by {\tt th}.
Then, line 16 generates a {\tt GLT\_tasklet} that contains the function pointer ({\tt L4}), the function arguments 
({\tt L4args}) and the tasklet handler. This Tasklet will be responsible for executing the corresponding iteration space of {\tt jr}, including
Loop~5 and the micro-kernel(s).
Line 19 allows the current thread to yield and start executing pending work-units (Tasklets). 
Finally, line 22 checks the Tasklet status to ensure that the work has been completed (synchronization point).

In Section~\ref{sec:experiments}, 
we evaluate the LWT solutions based on GLTO vs GLTO+GLT, and we compare the performance compared with a
conventional OpenMP runtime using the DMF algorithms as the target case study.

%% file: s6-experiments.tex
\section{Performance Evaluation}
\label{sec:experiments}

\subsection{Experimental setup}

All the experiments in this paper 
were performed in double precision real arithmetic,
on a server equipped with an 8-core Intel Xeon E5-2630 v3 (``Haswell'') processor, running at 2.4 GHz, 
and 64 Gbytes of DDR4 RAM.
The codes were compiled with Intel {\tt icc} 17.0.1 or GNU {\tt gcc} 6.3.0. 
The LWT implementation is that in Argobots.\footnote{Version from October 2017. Available online
at \url{http://www.argobots.org}.}
(Unless explicitly stated otherwise, we will use Intel's compiler and OpenMP runtime.)
The instance of BLAS is a modified version of BLIS 0.1.8, to accommodate malleability,
where the cache configuration parameters were set to
$n_c$ = 4032,
$k_c$ = 256,
$m_c$ = 72,
$n_r$ = 6, and
$m_r$ = 8. These values are optimal for the Intel Haswell architecture.

The matrices employed in the study are all square of order $n$, with random entries following a uniform distribution.
(The specific values can only have a mild impact on the execution time of LUpp, because of the different permutation sequences that they produce.)
The algorithmic block size for all algorithms was set to $\bs=192$. 
This specific value of $\bs$ is not particularly biased to favor any of the algorithms/implementations
and avoids a very time-consuming optimization of this parameter for space range of tuples DMF/problem dimension/implementation.

In the following two subsections, we employ LUpp to compare the distinct behavior of Intel's and GNU's runtimes 
when dealing with nested parallelism;
and the performance differences when using GLTO or GLT to parallelize BLAS. After identifying the best options
with these initial analyses, in the subsequent subsection 
we perform a global comparison using three DMFs: LUpp, the QR factorization (QR), and a routine
for the reduction to band form that is utilized in the computation of the SVD. 
These DMFs are representative of many linear algebra codes in LAPACK.

\subsection{Conventional OS threads: GNU vs Intel}
\label{subsec:intelvsgnu}

GNU and Intel have different policies to deal with nested parallelism that may produce relevant consequences on performance. 
In principle, upon encountering the first (outer) parallel
region, say {\sf OR} (for outer region), 
both runtimes ``spawn'' the requested number of threads. For each thread
hitting the second (inner) region, say {\sf IR1} (inner region-1), 
they will next  ``spawn'' as many threads as requested in the corresponding directive. 
The differences appear when, after completing
the execution of {\sf IR1}, a new inner region {\sf IR2} is encountered. 
In this scenario, GNU's runtime will set the threads that executed 
{\sf IR1} to idle, and a new team of threads will be spawned and put
in control of executing
{\sf IR2}. 
Intel's runtime behavior differs from this in that it re-utilizes the team that executed
{\sf IR1} for
{\sf IR2} (plus/minus the differences in the number of threads requested by the two inner regions).
This discussion is important because, in our parallelization of the DMFs, this is exactly the scenario that occurs:
{\sf OR} is the region in the DMF algorithm that employs the {\tt parallel sections} directive, while
{\sf IR1},
{\sf IR2},
{\sf IR3},\ldots
correspond to each one of regions annotated with the 
{\tt parallel for} directives that are encountered in successive iterations of Loop~3 for the BLAS. It is thus easy to
infer that, under these circumstances, GNU will produce considerable oversubscription, 
due to the overhead of creating new teams even if the threads are set to a passive mode after no longer needed 
(or even worse if they actively cycle in a busy-wait).

With Intel, a mild risk of oversubscription still appears with the version of the DMF algorithm that employs a malleable BLAS.
In this case, the thread that completes the execution of the panel factorization, upon execution of this part, is set to
idle; and the next time the {\tt parallel for} inside Loop~3 of the BLAS is encountered, a new thread becomes part of the
team executing the panel update. The outcome is that now we have one thread waiting for the synchronization at the end
of the {\tt parallel sections} and {\tt tMM=t} threads executing the trailing update, where {\tt t} denotes the number
of cores. Fortunately, we can avoid the negative consequences in this case by controlling the behavior of the idle
thread via Intel's environment variables, as we describe next.

\begin{figure}[htb]
\centering
\includegraphics[width=0.6\textwidth]{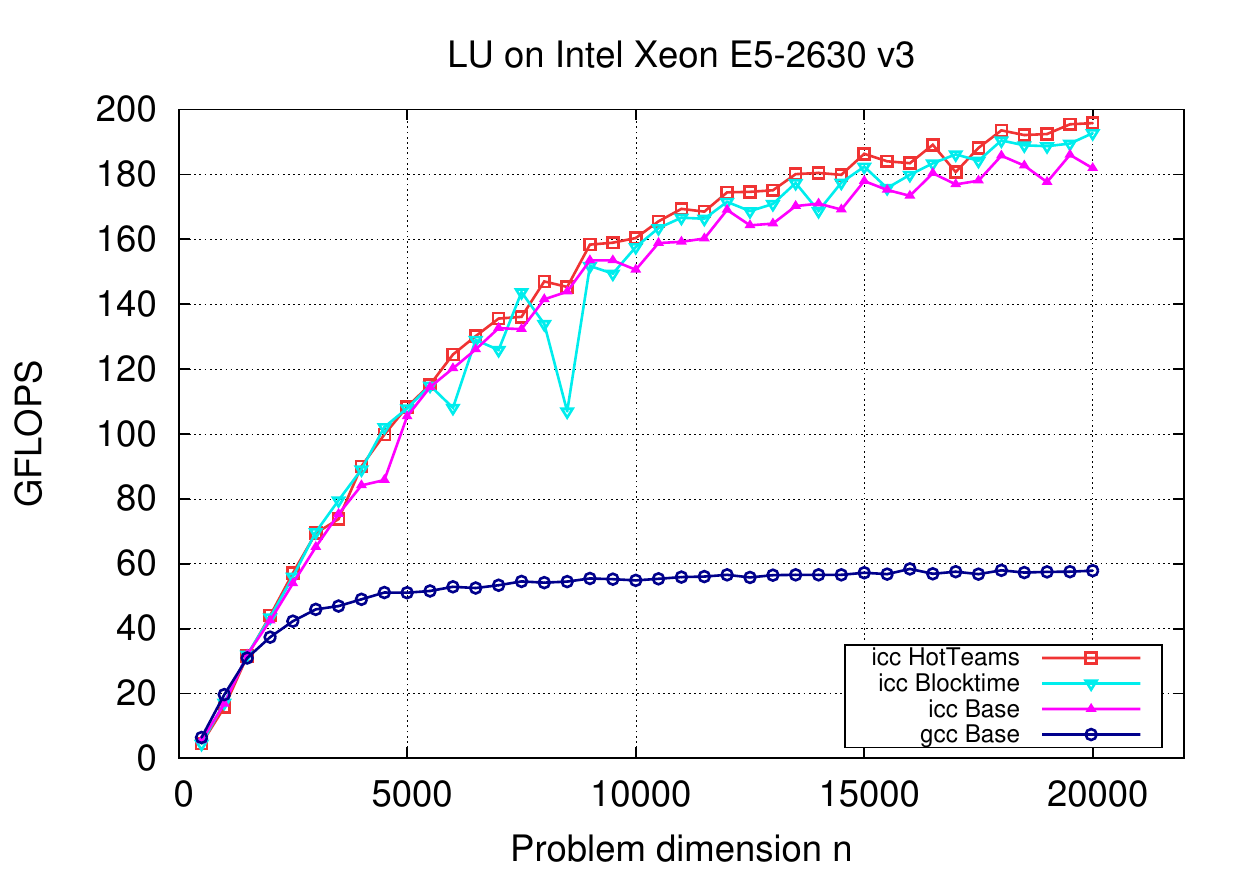}
\caption{Performance of LUpp using the conventional OpenMP runtimes on 8 cores of an Intel Xeon E5-2630 v3.}
\label{fig:gcc_vs_icc}
\end{figure}

The experiments in this subsection aim to illustrate these effects. Concretely, Figure~\ref{fig:gcc_vs_icc}
compares the performance of both conventional runtimes for the LUpp codes (with static look-ahead in all cases), 
and shows the impact of their
mechanisms for thread management in performance. For Intel's runtime, we also provide a more detailed
inspection using several fine-grained optimization strategies enforced via environment variables. 
Each line of the plot corresponds to a different combination of runtime-environment variables as follows:
\begin{description}
  \item[\sf Base:] Basic configuration for both runtimes. 
Nested parallelism is explicitly enabled by setting
{\tt OMP\_NESTED=true} and {\tt OMP\_MAX\_LEVELS=2}.
The waiting policy for idle threads is explicitly enforced to be passive for both runtimes via the initialization
{\tt OMP\_WAIT\_POLICY=passive}. This environment variable defines whether threads 
spin (active policy) or sleep (passive policy) while they are waiting.

  \item[\sf Blocktime:] Only available for Intel's runtime. When using a passive waiting policy, we leverage
variable {\tt KMP\_BLOCKTIME} to fix the time that a thread should wait after
completing the execution of a parallel region before sleeping. In our case, we have empirically
determined an optimal waiting time of 1 ms. (In comparison, the default value is 200 ms.)

  \item[\sf HotTeams:] Only available for Intel's runtime. {\em Hot teams} is an extension of 
OpenMP supported by the Intel runtime that specifies the runtime behavior when the number of threads in a team is reduced.
Specifically, when the {\em hot teams} are active, 
extra threads are kept in the team in reserve, for faster re-use in subsequent parallel regions,
potentially reducing the overhead 
associated with a full start/stop procedure. This functionality
by setting {\tt KMP\_HOT\_TEAMS\_MODE=1} and {\tt KMP\_HOT\_TEAMS\_MAX\_LEVEL=2}.

\end{description}

The analysis of performance in Figure~\ref{fig:gcc_vs_icc} exposes the differences between 
{\sf Base} configurations of the Intel's and GNU's runtimes, mainly derived from the distinct policies
in thread re-use between the two runtimes, and the consequent oversubscription problem described above. 
For Intel's runtime, the explicit introduction of a passive wait policy ({\sf Base} line)
yields a substantial performance boost compared with GNU; and additional performance gains are derived from 
the use of an optimal block time value, and {\em hot teams} (lines labeled with {\sf Blocktime} and {\sf HotTeams}, 
respectively).

\subsection{LWT in Argobots: GLTO vs GLTO+GLT}
\label{subsec:gltovsglt}

\begin{figure}[htb]
\centering
\includegraphics[width=0.6\textwidth]{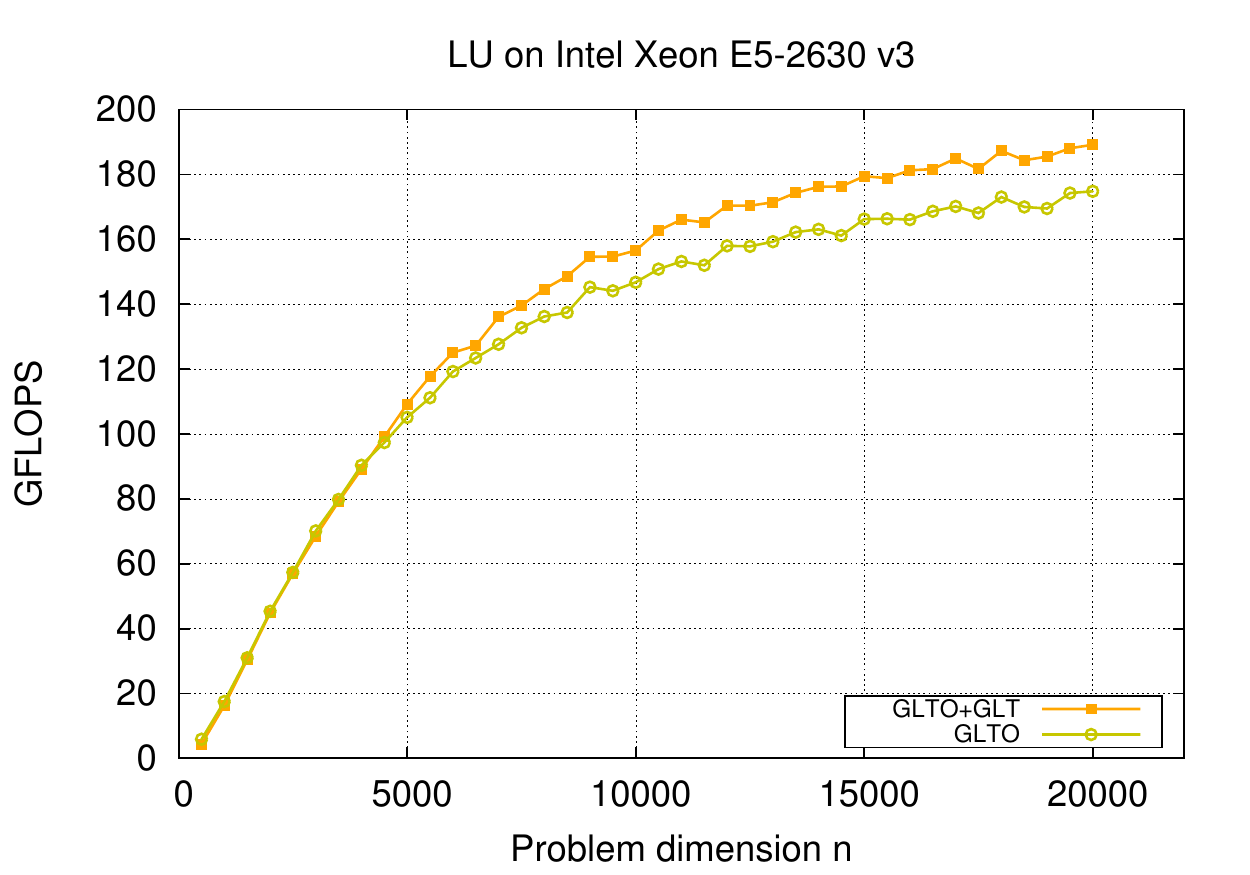}
\caption{Performance of LUpp using the LWT in Argobots on 8 cores of an Intel Xeon E5-2630 v3.}
\label{fig:gltovsglt}
\end{figure}

Figure~\ref{fig:gltovsglt}
compares the performance of the LUpp codes (with static look-ahead), 
using the two LWT solutions described in subsection~\ref{sec:LWT}. Here we remind that 
the simplest variant utilizes GLTO's OpenMP-API
on top of Argobot's runtime (line labeled as {\sf GLTO} in the plot) while the most
sophisticated one, in addition, employs Tasklets to parallelize the BLAS
(line {\sf GLTO+GLT}).
This experiment show that using Tasklets compensates the additional efforts of developing this specific implementation
of the BLAS. This is especially the case, 
as this development is a one-time effort that, once completed, can be seamlessly leveraged multiple times by the users of this 
specialized instance of the library.

\subsection{Global comparison}

The final analysis in this paper compares the five parallel algorithms/implementations listed next.
Unless otherwise stated, they all employ Intel's OpenMP runtime.
\begin{itemize}
\item {\sf MTB}: Conventional approach that extracts parallelism in the reference DMF routines (without look-ahead) by simply linking them with a multi-threaded instance of BLAS.
\item {\sf RTM}: Runtime-assisted parallelization that decomposes the trailing update into multiple tasks and
simultaneously executes independent tasks in different cores. 
Most of the tasks correspond to BLAS kernels which are executed using a serial
(i.e., single-threaded) instance of this library.
The tasks are identified using the OpenMP 4.5 {\tt task} directive and dependencies are specified via representants for the
blocks and the proper {\tt in}/{\tt out} clauses.
\item {\sf LA}:  DMF algorithm that integrates a static look-ahead and exploits NMP with task-parallelism extracted from the 
      loop-body of the factorization and loop-parallelism from the multi-threaded BLAS.
\item {\sf LA\_MB\_S} and {\sf LA\_MB\_G}: Analogous to {\sf LA} but linked with an MTL multi-threaded version of BLAS.
      The first implementation (with the suffix ``{\sf \_S}'') employs Intel's OpenMP runtime, with the environment variables set as determined in the
      study in subsection~\ref{subsec:intelvsgnu}. The second one (suffix ``{\sf \_G}'') employs GLTO+GLT
and Argobot's runtime, as derived to be the best option from the experiment in subsection~\ref{subsec:gltovsglt}.
\end{itemize}

For this study, we use leverage the following three DMFs:
\begin{itemize}
\item LUpp: The LU factorization with partial pivoting as utilized and described earlier in this work; see subsection~\ref{subsec:MTBvsRTM}.
\item QR: The QR factorization via Householder transformations. The reference implementation is a direct translation into C of routine {\sc geqrf} in LAPACK.
      The version with static look-ahead is obtained from this code by re-organizing the operations as explained for the generic DMF earlier
      in the paper. 
The runtime-assisted parallelization operates differently, in order to expose a higher degree of parallelism, 
      but due to the numerical stability of orthogonal transformations, produces the same result.
In particular, {\sf RTM}
      divides the panel and trailing submatrix 
      into square blocks, using the same approach proposed in~\cite{Buttari200938,Quintana-Orti:2009:PMA:1527286.1527288}, and derived
      from the incremental QR factorization in~\cite{Gunter:2005:POC}.
\item SVD: The reduction to compact band form for the (first stage of the) computation of the SVD, as described in~\cite{GROER1999969,DBLP:journals/corr/abs-1709-00302}. 
      This is a right-looking routine that,
      at each iteration, computes two panel factorizations, using Householder transformations respectively applied from the left- and right-hand side of the matrix.
      These transformations are next applied to update the trailing parts of the matrix via efficient BLAS-3 kernels.
      The variants that allow the introduction of static look-ahead were presented in~\cite{DBLP:journals/corr/abs-1709-00302}.
      No runtime version exist at present for this factorization~\cite{DBLP:journals/corr/abs-1709-00302}.
\end{itemize}
The results are compared in terms of GFLOPS, using the standard flop counts
for LUpp ($2n^3/3$) and QR ($4n^3/3$). For the SVD reduction routine, we employ the theoretical flop count
of $8n^3/3$ for the full reduction to bidiagonal form. 
However, the actual number of flops depends on the relation between the actual target bandwidth $w$
and the problem dimension. In these experiments, $w$ was set to 384.
For the SVD, this performance ratio allows a fair comparison between the different algorithms
as the GFLOPS can still be viewed as an scaled metric (for the inverse of) time.

\begin{figure}[t]
\centering
\includegraphics[width=0.6\textwidth]{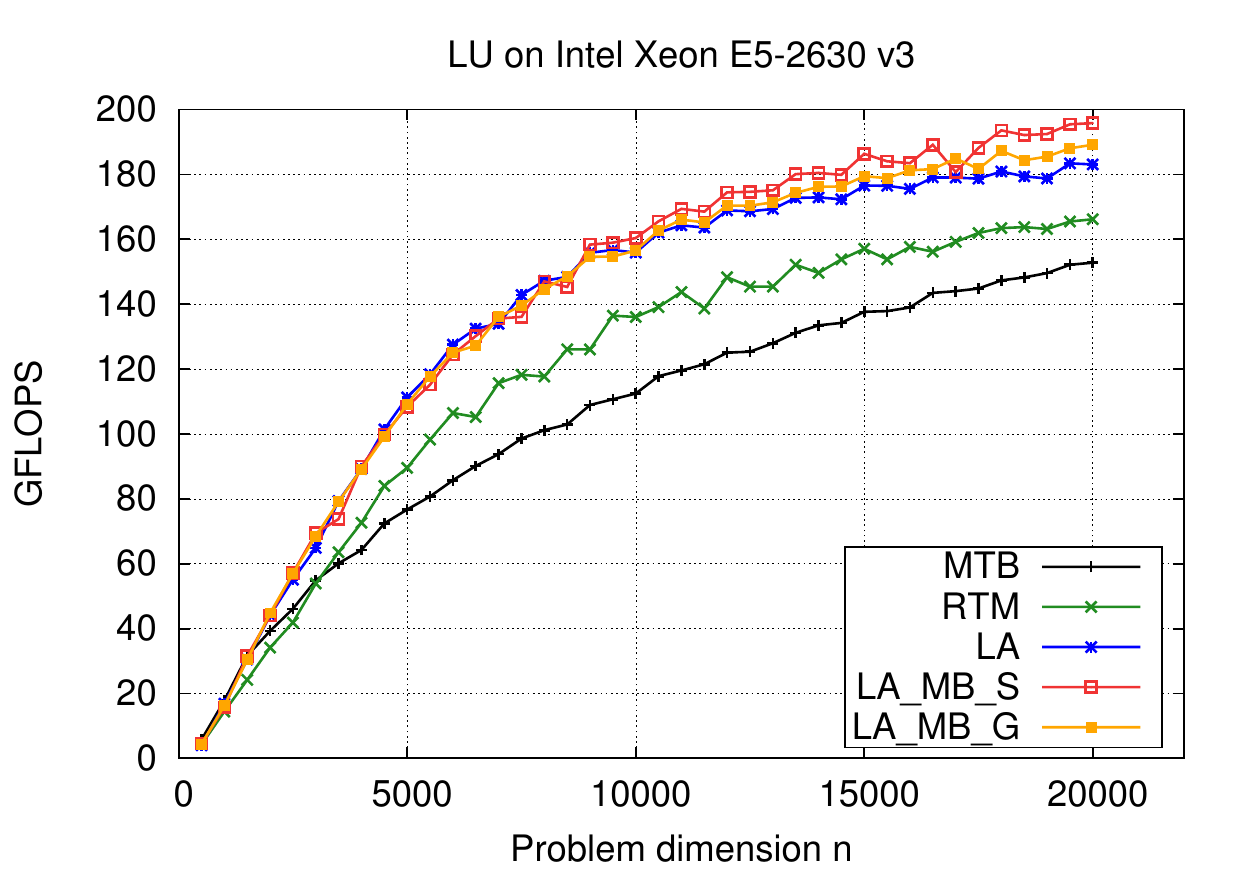}\\
\caption{Performance of LUpp on 8 cores of an Intel Xeon E5-2630 v3.}
\label{fig:comparison_LU}
\end{figure}

\begin{figure}[t]
\centering
\includegraphics[width=0.6\textwidth]{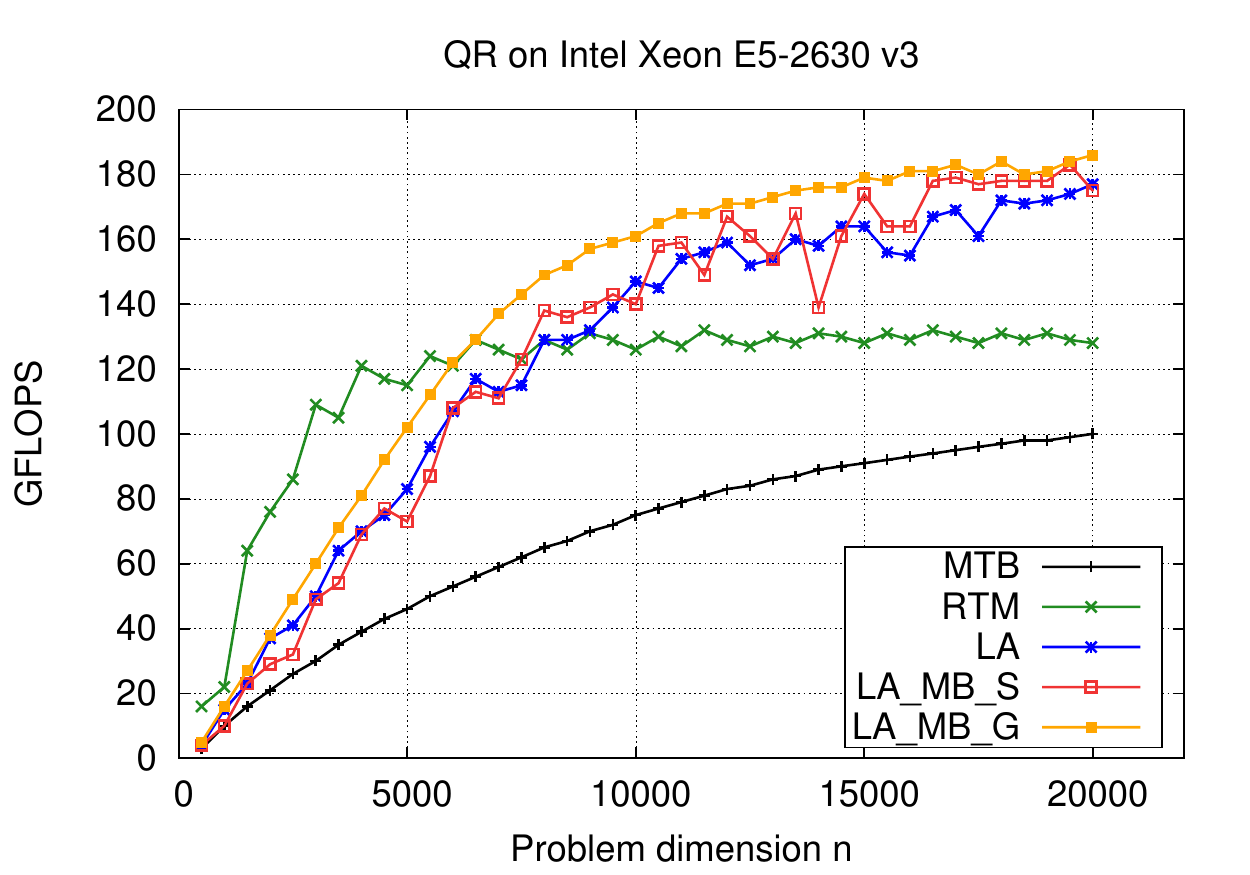}\\
\caption{Performance of QR on 8 cores of an Intel Xeon E5-2630 v3.}
\label{fig:comparison_QR}
\end{figure}

\begin{figure}[t]
\centering
\includegraphics[width=0.6\textwidth]{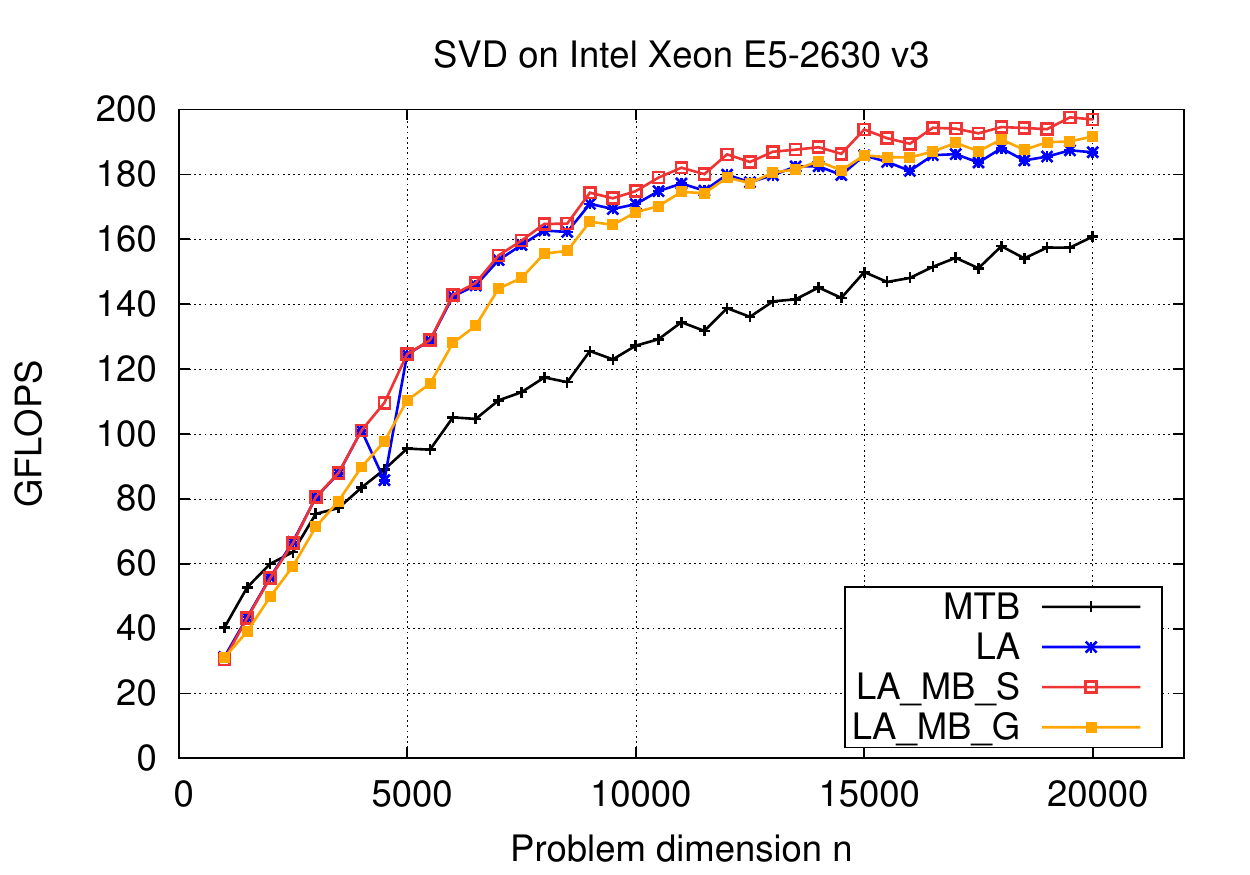}\\
\caption{Performance of SVD on 8 cores of an Intel Xeon E5-2630 v3.}
\label{fig:comparison_SVD}
\end{figure}

Figures~\ref{fig:comparison_LU}--\ref{fig:comparison_SVD} 
compare the performance of the distinct algorithms for the three DMFs, 
using square matrices of growing dimensions from 500 till 20,000 in steps of 500. 
These experiments offer some important insights:
\begin{itemize}
\item The basic algorithm ({\sf MTB}), corresponding to the reference implementation without look-ahead, which extracts 
      all parallelism from the BLAS, cannot compete with the other variants. 
      The reason for this is the low performance of the panel factorization, which stands in the critical path of the 
      algorithm, and results in a serious bottleneck for the global performance of the algorithm.
      (Decreasing drastically the panel width, i.e., the algorithmic block size \bs, is not an option because the
      trailing update then becomes a memory-bound kernel, delivering low performance and poor parallel scalability.)
\item The algorithm enhanced with a static look-ahead ({\sf LA}) partially eliminates the problem of the panel factorization
      by overlapping, at each iteration, the execution of this operation with that of the highly-parallel trailing update.
      Only for the smallest problem sizes, the panel factorization is too expensive compared with the trailing update,
      and the cost of the panel operation cannot be completely hidden.
      (However, as stated earlier, this an be partially tackled via early termination~\cite{catalan17}.)
\item As the problem size grows, employing a malleable instance of BLAS (as in versions
      {\sf LA\_MB\_S} and
      {\sf LA\_MB\_G}) squeezes around 5--20 additional GFLOPS (depending on the DMF and problem dimension)
      with respect to the version with look-ahead
      that employs the regular implementation of BLAS. This comes from the thread performing the panel factorization
      jumping into the trailing update as soon as it is done with the former operation. As it was expected,
      this occurs for the largest problems, as in those cases the cost of the trailing update dominates over the
      panel factorization. Furthermore, the theoretical performance advantage that could be expected is 8/7 (from using
      7 threads in the trailing update to having 8), which is about 14\% at most, in the theoretical assumption 
      that the panel factorization has no cost. This represents about 25~extra GFLOPS for a performance
      rate of 180~GFLOPS.
\item The runtime-based parallelization ({\sf RTM}) is clearly outperformed by the algorithms that integrate a static look-ahead
      for LUpp and all problem dimensions. This is a consequence of the excessive fragmentation into fine-grain kernels
      and the overhead associated with these conditions. The scenario though is different for QR. There {\sf RTM} is the best
      option for small problem sizes. The reason is that the algorithm for this factorization performs a more aggressive 
      division of the factorization into fine-grain tasks, which in this case pay offs for this range of problems. 
      Unfortunately, the same approach cannot be applied to LUpp without abandoning the standard partial pivoting 
      and, therefore, changing the numerics of the algorithm.
\end{itemize}

%% file: s7-remarks.tex
\section{Concluding Remarks}
\label{sec:remarks}

We have addressed the parallelization of a general framework that accommodates a relevant number of dense linear
algebra operations, including the major dense matrix factorizations (LU, Cholesky, QR and LDL$^{{\rm T}}$), matrix
inversion via Gauss-Jordan elimination, and the initial decomposition 
in two-stage algorithms for the reduction to compact band forms 
for the solution of symmetric eigenvalue problems and the computation of the SVD.
Our work describes these algorithms with a high level of abstraction, hiding some implementation details, 
an employs a high-level parallel programming API such as OpenMP 
to provide enough information in order to obtain a practical high-performance parallel code for multicore processors.
The key factors to the success of this approach are:
\begin{itemize}
\item The exploitation of task-parallelism in combination with a static look-ahead strategy
      explicitly embedded in the code that hides the latency of the panel factorization. 
\item The integration of a malleable, multi-threaded instance of the BLAS that realizes the major part of the flops 
      and ensures that the threads/cores involved in these operations efficiently share the memory resources causing
      little overhead.
\item The use of Intel's OpenMP runtime, with the proper setting of several environment variables in order to prevent 
      oversubscription problems when exploiting nested parallelism or, alternatively, the support from a LWT-runtime such
      as Argobots. 
\end{itemize}
Our approach shows very competitive results, in general outperforming other parallelization strategies for DMFs,
for problem dimensions that are large enough with respect to the number of cores.

Overall, we recognize that
current development efforts in the DLA-domain are pointing in the direction of introducing dynamic scheduling via a runtime,
taking away the burden of optimization off the user while still providing high performance across different systems.
In comparison, when applied with care, one could naturally expect that
a manual distribution of the workload among the processor cores outperforms dynamic scheduling, at the cost of
a more complex coding effort.
This work aims to show that, given the right level of abstraction, modifying a DMF routine to manually introduce a static
look-ahead, and parallelizing the outcome via the appropriate runtime, is a simple task.

%% file: sn-acks.tex
This work was supported by the CICYT projects TIN2014-53495-R 
and TIN2017-82972-R of the MINECO and FEDER, and the H2020 EU FETHPC Project 671602 ``INTERTWinE''. 
Sandra Catal\'an was supported during part of this time
by the FPU program of the {\em Ministerio de Educaci\'on, Cultura y Deporte}.
Adri\'an Castell\'o was supported by the ValI+D 2015 FPI program of the {\em Generalitat Valenciana}.

%% file: paper.bbl
\begin{thebibliography}{10}

\bibitem{lapack}
Edward Anderson, Zhaojun Bai, L.~Susan Blackford, James Demmel, Jack~J.
  Dongarra, Jeremy~Du Croz, Sven Hammarling, Anne Greenbaum, Alan McKenney, and
  Danny~C. Sorensen.
\newblock {\em {LAPACK} Users' guide}.
\newblock SIAM, 3rd edition, 1999.

\bibitem{BadiaHLPQQ09}
Rosa~M. Badia, Jose~R. Herrero, Jesus Labarta, Jose~M. P\'erez, Enrique~S.
  Quintana-Ort\'{\i}, and Gregorio Quintana-Ort\'{\i}.
\newblock Parallelizing dense and banded linear algebra libraries using
  {SMPSs}.
\newblock {\em Conc. and Comp.: Pract. and Exper.}, 21:2438--2456, 2009.

\bibitem{FLAME:Recipe}
Paolo Bientinesi, John~A. Gunnels, Margaret~E. Myers, E.~S. Quintana-Ort\'{\i},
  and Robert~A. van~de Geijn.
\newblock The science of deriving dense linear algebra algorithms.
\newblock {\em ACM Trans. Math. Softw.}, 31(1):1--26, 2005.

\bibitem{Bischof:2000:AST}
Christian~H. Bischof, Bruno Lang, and Xiaobai Sun.
\newblock {Algorithm 807}: {The SBR Toolbox}---software for successive band
  reduction.
\newblock {\em ACM Trans. Math. Soft.}, 26(4):602--616, 2000.

\bibitem{Buttari200938}
Alfredo Buttari, Julien Langou, Jakub Kurzak, and Jack Dongarra.
\newblock A class of parallel tiled linear algebra algorithms for multicore
  architectures.
\newblock {\em Parallel Computing}, 35(1):38 -- 53, 2009.

\bibitem{lwthlpm}
Adrián Castelló, Rafael Mayo, Kevin Sala, Vicenç Beltran, Pavan Balaji, and
  Antonio~J. Peña.
\newblock On the adequacy of lightweight thread approaches for high-level
  parallel programming models.
\newblock {\em Future Generation Computer Systems}, 84:22 -- 31, 2018.

\bibitem{cluster16}
Adri{\'a}n Castell{\'o}, Antonio~J. {Pe\~na}, Sangmin Seo, Rafael Mayo, Pavan
  Balaji, and Enrique~S. Quintana-Ort\'i.
\newblock A review of lightweight thread approaches for high performance
  computing.
\newblock In {\em Proceedings of the IEEE International Conference on Cluster
  Computing}, Taipei, Taiwan, September 2016.

\bibitem{GLTAPI}
Adri{\'a}n Castell{\'o}, Sangmin Seo, Rafael Mayo, Pavan Balaji, Enrique~S.
  Quintana-Ort\'i, and Antonio~J. {Pe\~na}.
\newblock {GLT: A unified API for lightweight thread libraries}.
\newblock In {\em Proceedings of the IEEE International European Conference on
  Parallel and Distributed Computing}, Santiago de Compostela, Spain, August
  2017.

\bibitem{GLTO}
Adri{\'a}n Castell{\'o}, Sangmin Seo, Rafael Mayo, Pavan Balaji, Enrique~S.
  Quintana-Ort\'i, and Antonio~J. {Pe\~na}.
\newblock {GLTO}: On the adequacy of lightweight thread approaches for {OpenMP}
  implementations.
\newblock In {\em Proceedings of the International Conference on Parallel
  Processing}, Bristol, UK, August 2017.

\bibitem{catalan17}
Sandra Catal{\'{a}}n, Jos{\'{e}}~R. Herrero, Enrique~S. Quintana{-}Ort{\'{\i}},
  Rafael Rodr{\'{\i}}guez{-}S{\'{a}}nchez, and Robert~A. van~de Geijn.
\newblock A case for malleable thread-level linear algebra libraries: The {LU}
  factorization with partial pivoting.
\newblock {\em CoRR}, abs/1611.06365, 2016.

\bibitem{Catalan2016}
Sandra Catal{\'a}n, Francisco~D. Igual, Rafael Mayo, Rafael
  Rodr{\'i}guez-S{\'a}nchez, and Enrique~S. Quintana-Ort{\'i}.
\newblock Architecture-aware configuration and scheduling of matrix
  multiplication on asymmetric multicore processors.
\newblock {\em Cluster Computing}, 19(3):1037--1051, 2016.

\bibitem{chamaleonweb}
{Chameleon} project.
\newblock \url{http:https://project.inria.fr/chameleon/}.

\bibitem{Dem97}
J.~Demmel.
\newblock {\em Applied Numerical Linear Algebra}.
\newblock Society for Industrial and Applied Mathematics, 1997.

\bibitem{BLAS3}
Jack~J. Dongarra, Jeremy Du~Croz, Sven Hammarling, and Iain Duff.
\newblock A set of level 3 basic linear algebra subprograms.
\newblock {\em ACM Trans. Math. Softw.}, 16(1):1--17, March 1990.

\bibitem{flameweb}
{FLAME} project home page.
\newblock \url{http://www.cs.utexas.edu/users/flame/}.

\bibitem{GVL3}
Gene~H. Golub and Charles F.~Van Loan.
\newblock {\em Matrix Computations}.
\newblock The Johns Hopkins University Press, Baltimore, 3rd edition, 1996.

\bibitem{Goto:2008:AHM:1356052.1356053}
Kazushige Goto and Robert A. van~de Geijn.
\newblock Anatomy of high-performance matrix multiplication.
\newblock {\em ACM Trans. Math. Softw.}, 34(3):12:1--12:25, May 2008.

\bibitem{Goto:2008:HPI}
Kazushige Goto and Robert van~de Geijn.
\newblock High performance implementation of the level-3 {BLAS}.
\newblock {\em {ACM} Transactions on Mathematical Software}, 35(1):4:1--4:14,
  July 2008.

\bibitem{GROER1999969}
Benedikt Grosser and Bruno Lang.
\newblock Efficient parallel reduction to bidiagonal form.
\newblock {\em Parallel Computing}, 25(8):969 -- 986, 1999.

\bibitem{Gunter:2005:POC}
Brian~C. Gunter and Robert~A. van~de Geijn.
\newblock Parallel out-of-core computation and updating the {QR} factorization.
\newblock {\em ACM Trans. Math. Soft.}, 31(1):60--78, March 2005.

\bibitem{ESSL}
IBM.
\newblock {E}ngineering and {S}cientific {S}ubroutine {L}ibrary.
\newblock \url{http://www-03.ibm.com/systems/power/software/essl/}, 2015.

\bibitem{MKL}
Intel.
\newblock {M}ath {K}ernel {L}ibrary.
\newblock \url{https://software.intel.com/en-us/intel-mkl}, 2015.

\bibitem{ompssweb}
{OmpSs} project home page.
\newblock \url{http://pm.bsc.es/ompss}.

\bibitem{OpenBLAS}
\url{http://www.openblas.net}, 2015.

\bibitem{openmp}
The {OpenMP API} specification for parallel programming.
\newblock \url{http://www.openmp.org}, 2017.

\bibitem{plasmaweb}
{PLASMA} project home page.
\newblock \url{http://icl.cs.utk.edu/plasma}.

\bibitem{Quintana-Orti:2008}
E.~S. Quintana-Ort\'{i} and R.~A. van~de Geijn.
\newblock Updating an {LU} factorization with pivoting.
\newblock {\em ACM Trans. Math. Softw.}, 35(2):11:1--11:16, July 2008.

\bibitem{Quintana-Orti:2009:PMA:1527286.1527288}
Gregorio Quintana-Ort\'{\i}, Enrique~S. Quintana-Ort\'{\i}, Robert~A.
  {van~de~Geijn}, Field~G. {Van~Zee}, and Ernie Chan.
\newblock Programming matrix algorithms-by-blocks for thread-level parallelism.
\newblock {\em ACM Trans. Math. Softw.}, 36(3):14:1--14:26, 2009.

\bibitem{DBLP:journals/corr/abs-1709-00302}
Rafael Rodr{\'{\i}}guez{-}S{\'{a}}nchez, Sandra Catal{\'{a}}n, Jos{\'{e}}~R.
  Herrero, Enrique~S. Quintana{-}Ort{\'{\i}}, and Andr{\'{e}}s~E. Tom{\'{a}}s.
\newblock Two-sided reduction to compact band forms with look-ahead.
\newblock {\em CoRR}, abs/1709.00302, 2017.

\bibitem{argobots}
S.~Seo, A.~Amer, P.~Balaji, C.~Bordage, G.~Bosilca, A.~Brooks, P.~Carns,
  A.~Castell{\'o}, D.~Genet, T.~Herault, S.~Iwasaki, P.~Jindal, S.~Kale,
  S.~Krishnamoorthy, J.~Lifflander, H.~Lu, E.~Meneses, M.~Snir, Y.~Sun,
  K.~Taura, and P.~Beckman.
\newblock Argobots: A lightweight low-level threading and tasking framework.
\newblock {\em IEEE Transactions on Parallel and Distributed Systems},
  PP(99):1--1, 2017.

\bibitem{BLIS3}
Tyler~M. Smith, Robert van~de Geijn, Mikhail Smelyanskiy, Jeff~R. Hammond, and
  Field~G. Van~Zee.
\newblock Anatomy of high-performance many-threaded matrix multiplication.
\newblock In {\em Proc. IEEE 28th Int. Parallel and Distributed Processing
  Symp.}, IPDPS'14, pages 1049--1059, 2014.

\bibitem{starpuweb}
{StarPU} project.
\newblock \url{http://runtime.bordeaux.inria.fr/StarPU/}.

\bibitem{stein1992}
Dan Stein and Devang Shah.
\newblock Implementing lightweight threads.
\newblock In {\em USENIX Summer}, 1992.

\bibitem{Str98}
Peter Strazdins.
\newblock A comparison of lookahead and algorithmic blocking techniques for
  parallel matrix factorization.
\newblock Technical Report TR-CS-98-07, Department of Computer Science, The
  Australian National University, {Canberra} 0200 {ACT}, {Australia}, 1998.

\bibitem{BLIS1}
Field~G. {Van~Zee} and Robert~A. {van~de~Geijn}.
\newblock {BLIS}: A framework for rapidly instantiating {BLAS} functionality.
\newblock {\em ACM Trans. Math. Softw.}, 41(3):14:1--14:33, 2015.

\bibitem{atlas}
R.~Clint Whaley and Jack~J. Dongarra.
\newblock Automatically tuned linear algebra software.
\newblock In {\em Proceedings of SC'98}, 1998.

\bibitem{BLIS2}
Field G.~Van Zee, Tyler~M. Smith, Bryan Marker, Tze~Meng Low, Robert A. Van~De
  Geijn, Francisco~D. Igual, Mikhail Smelyanskiy, Xianyi Zhang, Michael
  Kistler, Vernon Austel, John~A. Gunnels, and Lee Killough.
\newblock The {BLIS} framework: Experiments in portability.
\newblock {\em ACM Trans. Math. Softw.}, 42(2):12:1--12:19, June 2016.

\end{thebibliography}
